\documentclass[aps,pra,reprint,groupedaddress,amsmath,amssymb]{revtex4-2}

\usepackage{graphicx}
\usepackage{dcolumn}
\usepackage{bm}

\begin{document}


\title{Intensity-dependent precision two-photon Doppler-free spectroscopy of Xe using narrow-bandwidth long-pulse deep-UV laser radiation}


\author{M. H. Rayment}
\affiliation{Institute of Molecular Physical Science, ETH Zürich, 8093 Zürich, Switzerland}

\author{R. Stech}
\affiliation{Institute of Molecular Physical Science, ETH Zürich, 8093 Zürich, Switzerland}

\author{F. Merkt}
\email[]{frederic.merkt@phys.chem.ethz.ch}
\affiliation{Institute of Molecular Physical Science, ETH Zürich, 8093 Zürich, Switzerland}


\date{\today}

\begin{abstract}
We report on a precision measurement of the $(5\mathrm{p})^{5}6\mathrm{p} \, [1/2]_{0}\leftarrow (5\mathrm{p})^{6} \, ^{1}\mathrm{S}_{0}$ transition wavenumber of Xe by Doppler-free two-photon spectroscopy using near-Fourier-transform-limited long pulses of UV-laser radiation. The measurements led to absolute uncertainties of 750~kHz in the two-photon transition wavenumbers for the five dominant isotopes of Xe, e.g., 80118.982918(27)~cm$^{-1}$ for $^{136}$Xe. These values were combined with other precision measurements in Xe to resolve an $\sim1~\mathrm{GHz}$ discrepancy between the ionization energy of Xe obtained in recent measurements from the $(6{\rm s})[3/2]_2$ metastable state [Herburger {\it et al.} Phys. Rev. A {\bf 109}, 032816 (2024)]
and the ionization energy listed in the NIST atomic database. The analysis indicates that the natural-abundance-weighted ionization energy of Xe should be revised by $-0.023$~cm$^{-1}$ to 97833.7641(20)~cm$^{-1}$.
Large fluctuations in the UV-laser pulse intensities were exploited to 
characterize intensity-dependent shifts of the observed two-photon transition frequencies. Shifts of up to $- 20$~MHz were observed and attributed to frequency shifts arising from chirps in the amplification and upconversion of the laser radiation rather than to the ac-Stark effect.
\end{abstract}


\maketitle
\newpage
\section{Introduction}

Doppler-free two-photon spectroscopy, invented 50 years ago \cite{biraben74a,levenson74a,haensch74a,biraben19a}, is one of the most powerful high-resolution spectroscopic methods and its applications include metrology (see, e.g., \cite{haensch75a,parthey11a}), precision measurements of the structure and dynamics of atoms and molecules (see, e.g., \cite{demtroeder11a,dickenson12a,zhao20a,lai21a,roth23a}) and trace-gas sensing (see, e.g., \cite{zhao20a,liu25a}). Its power results from the suppression of first-order Doppler broadening when the two-photon transition involves absorption of two photons of the same frequency but from exactly counterpropagating laser beams. Two-photon transitions in atoms and molecules are typically weak and high laser intensities are required to drive them. Their measurements are therefore prone to ac-Stark shifts and line broadening \cite{bloembergen76a,bergquist96a}, which need to be evaluated in experiments carried out at different intensities.

To achieve efficient two-photon excitation in the UV range, pulsed laser radiation is often used in combination with nonlinear frequency upconversion (second-harmonic generation or sum-frequency mixing) to generate the pulsed UV radiation. In such experiments, the resolution can be limited by the UV-laser pulse length, especially when Nd:YAG-pumped dye lasers and optical parametric oscillators are used, and frequency chirps need to be carefully measured and controlled \cite{eikema96a,eikema97a,bergeson98a,bergeson00a,hollenstein00a,brandi02a,seiler05a,paul05a,hannemann06a,liu09b,niu13a,cheng18a,beyer19a,herburger19a,hussels21a,hussels22a,hoelsch23a,heiss25a,eyler97a,white04a}.

We report here on the use of a long-pulse UV laser system \cite{seiler05a,paul05a} in a Doppler-free measurement of the $(5\mathrm{p})^{5}6\mathrm{p} \, [1/2]_{0}\leftarrow (5\mathrm{p})^{6} \, ^{1}\mathrm{S}_{0}$ two-photon transition of atomic Xe near 80118.9~cm$^{-1}$. Near-IR (NIR) pulses of adjustable duration up to 240~ns were produced by amplification, in Nd:YAG-pumped Ti:Sa bow-tie amplifiers, of seed pulses of adjustable length cut out of single-mode cw NIR laser radiation using an acousto-optic modulator. They were then converted to the UV by frequency tripling using two $\beta$-barium-borate crystals. Small shot-to-shot fluctuations in the Nd:YAG pump-laser and seed-laser pulses were strongly enhanced in the pulse amplification and nonlinear frequency upconversion from the NIR to the UV. We detail the use of a measurement technique for performing precision Doppler-free two-photon UV-laser spectroscopy with such a system which turns the very large fluctuations of the UV laser intensities into an advantage. The technique exploits single-shot measurements of the UV laser intensity and of Xe$^+$ ions generated by two-photon resonance-enhanced three-photon ionization ((2+1) REMPI) to record intensity-dependent frequency shifts. It enables the distinction between frequency shifts in the UV-laser frequency arising from frequency chirps induced by the amplification gain media and ac-Stark shifts of the two-photon transition frequencies.
Understanding and distinguishing different sources of frequency shifts is crucial for precision UV and VUV spectroscopic measurements that utilize narrow-bandwidth pulse-amplified laser radiation in high-resolution atomic and molecular spectroscopy~\cite{eikema96a,eikema97a,bergeson98a,bergeson00a,hollenstein00a,brandi02a,seiler05a,paul05a,hannemann06a,liu09b,niu13a,cheng18a,beyer19a,herburger19a,hussels21a,hussels22a,hoelsch23a,heiss25a}. 

The choice of the $(5\mathrm{p})^{5}6\mathrm{p} \, [1/2]_{0}\leftarrow (5\mathrm{p})^{6} \, ^{1}\mathrm{S}_{0}$ two-photon transition of atomic Xe for this measurements is motivated by its use in the generation of VUV radiation by resonance-enhanced four-wave mixing~\cite{yiu82a,page87a,cromwell89a,tunnermann93a,wellegehausen96a,paul05a} and by the need to clarify the origin of an $\sim1~\mathrm{GHz}$ discrepancy between a recent precision measurement of the first-ionization energy of Xe from the metastable $(5\mathrm{p}){^5} \, 6\mathrm{s} \, [3/2]_{2}$ state~\cite{herburger24a} and the recommended value of the first-ionization energy from the ground $(5\mathrm{p})^{6} \, ^{1}\mathrm{S}_{0}$ state~\cite{nist_asd_template,brandi01a,sukhorukov12a}.

The structure of the remainder of this article is as follows. In Sec.~\ref{Sec:Experiment} the apparatus and techniques used in the experiment are outlined. The results of the high-resolution (2+1) REMPI spectroscopy are presented in Sec.~\ref{Sec:Results}. Their implications in the context of the determination of first-ionization energy of Xe are discussed in Sec.~\ref{Sec:Discussion}. Conclusions are presented in Sec.~\ref{Sec:Conclusions}.

\section{Experiment}
\label{Sec:Experiment}

\begin{figure*}[tb!]
\includegraphics[width = \textwidth]{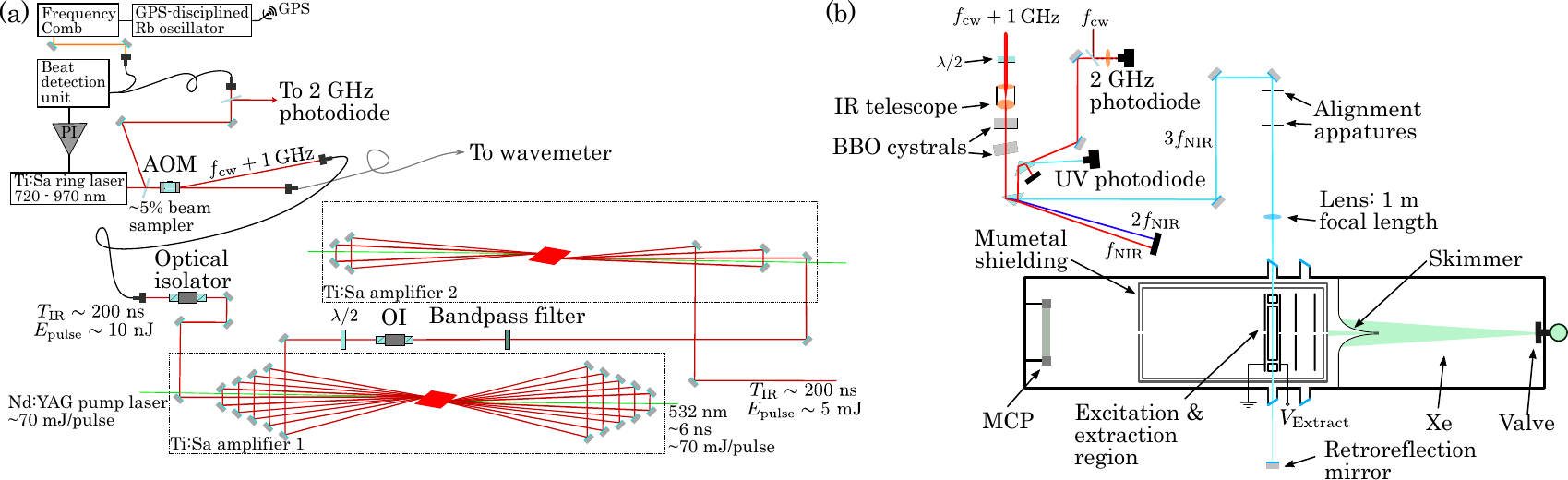}
\caption{Schematic diagram (not to scale) of the experimental setup showing (a)~the generation of the Fourier-transform-limited NIR laser pulses and (b)~the optical layout used to triple the NIR frequency with two BBO crystals and the vacuum chamber with the supersonic beam source, the photoexcitation region and the time-of-flight mass spectrometer. \label{Figure1}}
\end{figure*}

\begin{figure}[tb!]
\includegraphics[width = \columnwidth]{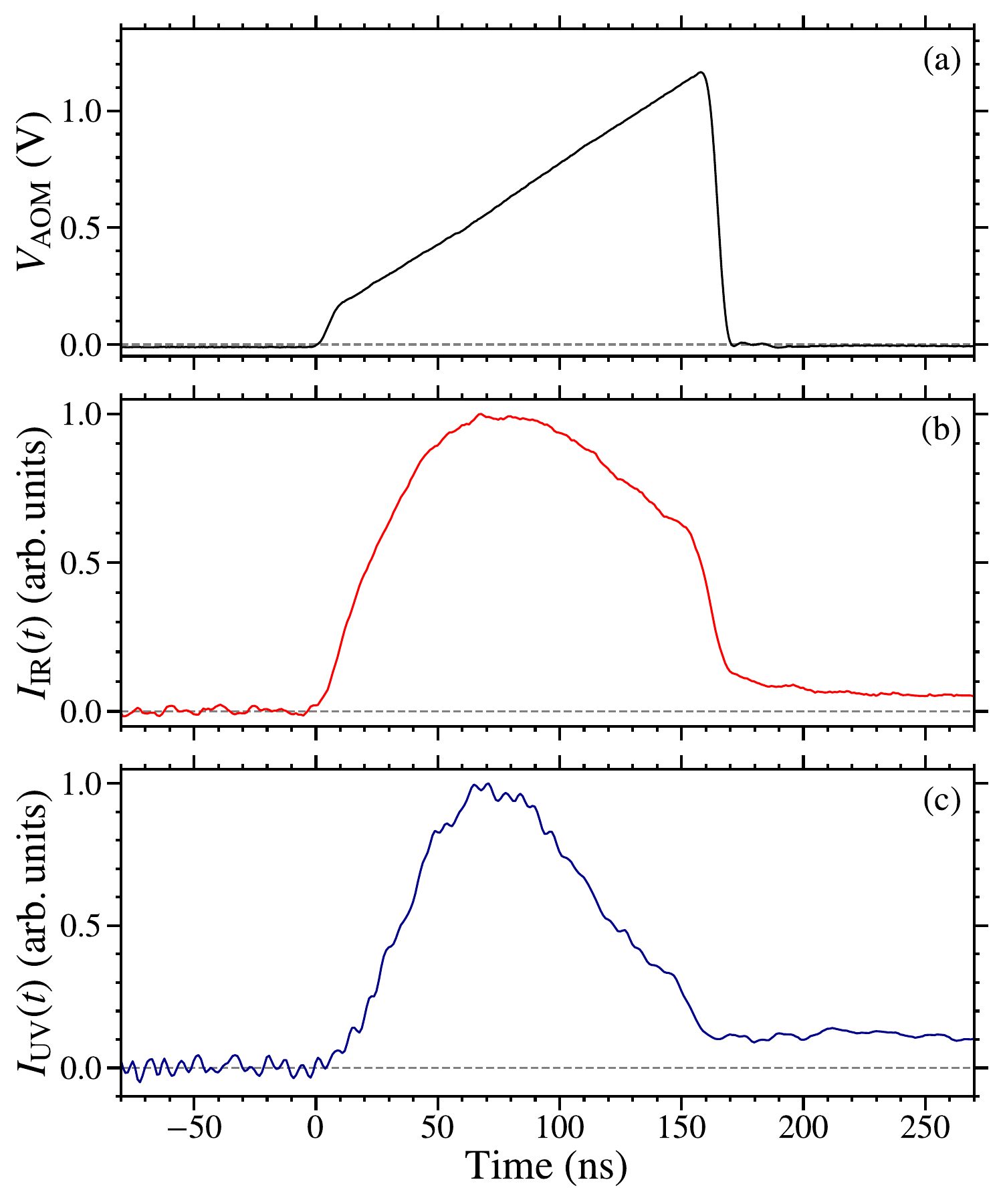}
\caption{Comparison of (a)~the pulse applied to the AOM to generate a pulse of coherent NIR radiation at frequency $f_{\mathrm{NIR}} \simeq f_{\mathrm{cw}} + f_{\mathrm{AOM}}$ out of the cw NIR-laser output ($f_{\mathrm{cw}}$) with (b)~the average of $\sim$2000 $160$-$\mathrm{ns}$-long NIR-output pulses after amplification with a pair of multipass bow-tie amplifiers, and (c)~the corresponding average UV output at $f_{\mathrm{UV}} \simeq 3f_{\mathrm{NIR}}$ after frequency upconversion with two BBO crystals. See text for details. 
\label{Figure2}}
\end{figure}

 A schematic diagram of the experimental setup is presented in Fig.~\ref{Figure1}. Figure~\ref{Figure1}(a) shows the setup used to generate narrow-bandwidth long-pulses of NIR laser radiation, which is similar to that described in Refs.~\cite{seiler05a,scheidegger22a}. Narrow-bandwidth continuous-wave (cw) NIR light at a wavelength of $\lambda_{\mathrm{cw}} \sim 748.8~\mathrm{nm}$ and a power of $\sim 700~\mathrm{mW}$ was generated using a titanium-doped sapphire (Ti:Sa) ring laser (Coherent 899) pumped with a Nd:YVO$_{4}$ cw diode laser (Coherent Verdi V8). A few percent of the cw NIR laser output was used to lock the Ti:Sa ring-laser frequency $f_{\mathrm{cw}}$ to the nearest-lying tooth of an optical frequency comb (Menlo Systems INC. FC1500-250-WG) referenced to a GPS-disciplined Rb-clock (Stanford Research Systems FS725). The rest of the cw NIR-laser output was passed through an acousto-optic modulator (AOM, Brimrose GPF-1000-500-800) operated in a pulsed mode at a repetition rate of 25 Hz and a driving frequency of $f_{\mathrm{AOM}}=1~\mathrm{GHz}$. The undiffracted zeroth-order output of the AOM, at frequency $f_{\mathrm{cw}}$, was coupled into a wavemeter (Highfinese WS7-60) with an absolute accuracy of $\sim 60~\mathrm{MHz}$, sufficient to unambiguously determine the comb tooth to which the cw NIR-laser was locked to, as described in Ref.~\cite{deiglmayr16a}.

NIR laser pulses with frequency $f_{\mathrm{NIR}} = f_{\mathrm{cw}} + 1~\mathrm{GHz}$, controlled length, and pulse energies of $ \sim 10~\mathrm{nJ}$ were generated in the first diffraction order of the cw-Ti:Sa ring-laser input beam by driving the AOM using rf-radiation pulses with envelopes $V_{\mathrm{AOM}}(t)$ controlled by an arbitrary waveform generator (Agilent AWG-33250A)~\cite{seiler05a}. The accuracy of the rf-driving frequency is better than $\sim1~\mathrm{kHz}$.
These ``seed" pulses were then amplified to pulse energies of $\sim 5~\mathrm{mJ}$ using a pair of Nd:YAG-pumped Ti:Sa  amplifiers in a bow-tie configuration~\cite{seiler05a} (see Fig.~\ref{Figure1}(a)).  In the experiments reported here, pulses with lengths $T_{\mathrm{IR}} \sim 160$ and $240~\mathrm{ns}$ were used with an adjustable time delay $t_{\mathrm{delay}}$ set to $0$, $40$, $60$, or $80~\mathrm{ns}$ with respect to the Nd:YAG laser pulses employed to pump the Ti:Sa amplifiers.

Long pulses of UV laser radiation ($\lambda_{\mathrm{UV}} \sim 249.6~\mathrm{nm}$) were produced by tripling the frequency of the NIR laser pulses using two successive $\beta$-barium-borate (BBO) crystals (see Fig.~\ref{Figure1}(b)).
Fig.~\ref{Figure2} illustrates the procedure, introduced in Ref.~\cite{seiler05a}, followed to control the shape of the UV laser pulses by adjusting the envelope of the rf pulses applied to the AOM. A linearly growing ramp (Fig.~\ref{Figure2}(a)) was used to compensate for the gradual depletion of the population inversion in the Ti:Sa-amplifier crystals resulting from the amplification and to achieve NIR (Fig.~\ref{Figure2}(b)) and UV (Fig.~\ref{Figure2}(c))  laser pulses with more symmetric pulse profiles, as measured with a battery-operated photodiode (Thorlabs DET10A2) with a nominal $1~\mathrm{ns}$ rise time.

\begin{figure}[tb!]
\includegraphics[width = \columnwidth]{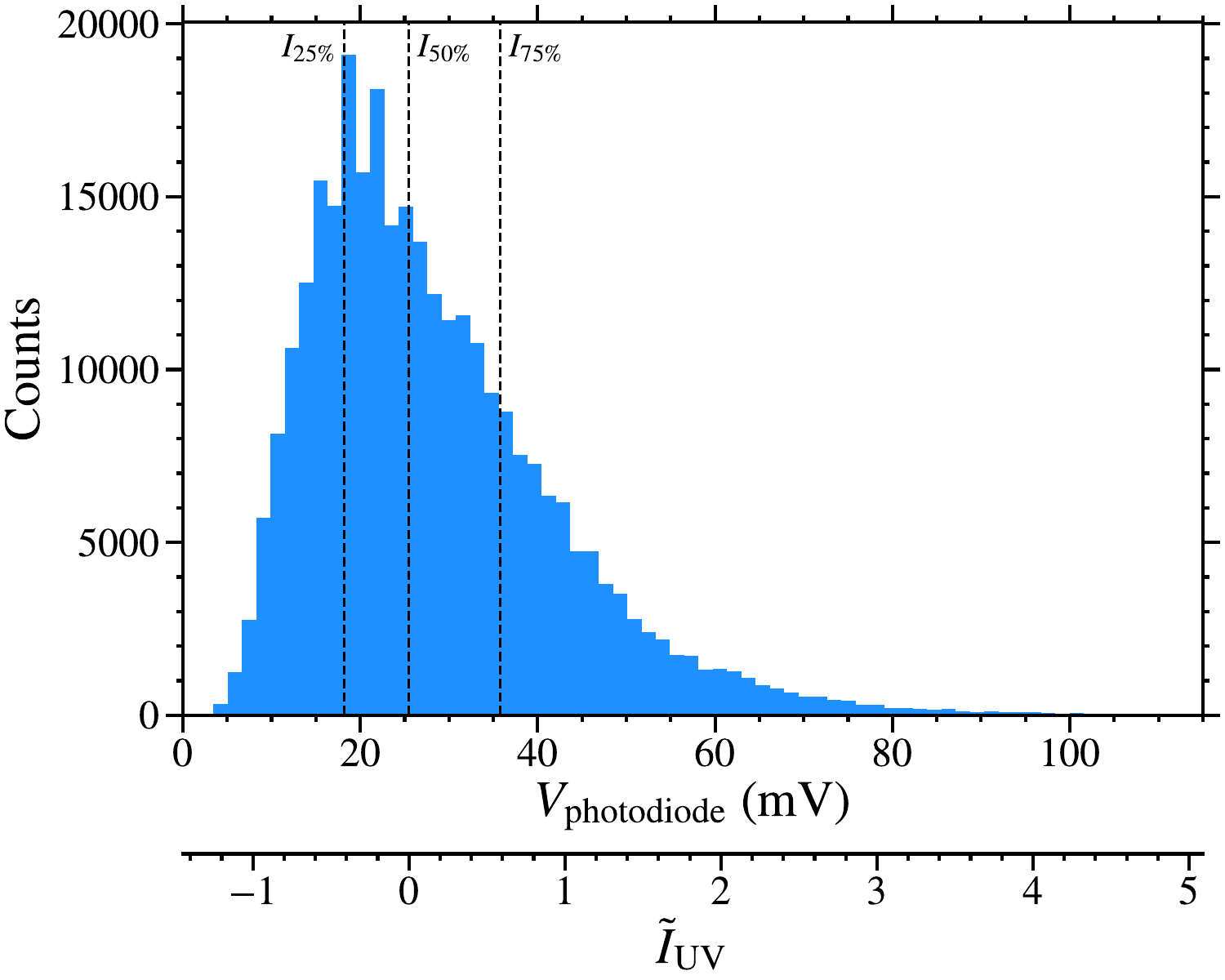} 
\caption{Histogram of the relative integrated single-shot photodiode signal $V_{\mathrm{photodiode}}$ used to monitor the pulsed UV laser intensities. The data corresponds to $\sim300000$ laser pulses generated by frequency upconversion of $240$-$\mathrm{ns}$-long NIR pulses. The upper and lower horizontal scales represent the photodiode signal in $\mathrm{mV}$ and the corresponding scaled intensity distribution $\tilde{I}_{\mathrm{UV}}$ as given by Eq.~(\ref{Eq:I_tilder}). The vertical dashed lines mark the $25~\%$, $50~\%$, and $75~\%$ percentiles to highlight the skewed nature of the distribution. \label{Figure3} }
\end{figure}

The large number of passes through the Ti:Sa crystals in the bow-tie amplifiers strongly enhances the effects of the fluctuations in the intensity of the NIR seed-laser pulses and of the Nd:YAG pump-laser pulses and causes large pulse-to-pulse variations in the pulsed-amplified NIR-laser output. These fluctuations are further amplified by the nonlinear frequency upconversion to the UV.
The shot-to-shot fluctuations of the UV-laser pulses are illustrated by the histogram depicted in Fig.~\ref{Figure3}, which shows a skewed distribution with a long tail towards high pulse energies. The asymmetry of the distribution of pulse energies is highlighted by the three dashed vertical lines, which indicate the position of the 25\%, 50\% (median) and 75\% percentiles. As a robust scaling of the asymmetric distribution, the lower horizontal scale of Fig.~\ref{Figure3} indicates the value of 
\begin{equation}
\label{Eq:I_tilder}
    \tilde{I}_{\mathrm{UV}} = \frac{I_{\mathrm{UV}} - I_{50\%}}{I_{75\%} - I_{25\%}},
\end{equation} 
where $I_{\mathrm{UV}}$ ($I_{\mathrm{UV}} \propto V_{\mathrm{photodiode}}$) represents the measured values of the relative intensity~\cite{dearmorim23a}. The unitless quantity $\tilde{I}_{\mathrm{UV}}$, which takes the value of 0 at the median, offers the advantage of a more universal description of intensity distributions recorded on different absolute scales. 

To record Doppler-free two-photon spectra, the UV laser radiation was focused using a $f_{\mathrm{lens}} = 1~\mathrm{m}$ focal-length lens onto a retroreflection mirror located on the other side of the vacuum chamber, $\sim50~\mathrm{cm}$ beyond the photoexcitation region.
A pair of alignment apertures (see Fig.~\ref{Figure1}(b)) of $\sim2.0~\mathrm{mm}$ diameter, located $185~\mathrm{cm}$ and $220~\mathrm{cm}$ from the retroreflection mirror were used to minimize potential deviations of the reflection angle from 180$^\circ$, which would cause systematic errors in the determination of the Doppler-free two-photon transition frequencies. This arrangement defines an upper bound $\theta_{\mathrm{max}} \simeq 0.54~\mathrm{mrad}$ for the misalignment angle between the forward-propagating  and retroreflected laser beams. 

In the experiments reported here, the sample was a skimmed pulsed supersonic beam of pure natural Xe gas generated with a Parker-General pulsed valve. The valve was operated at room temperature and a stagnation pressure of 2~bar, resulting in a beam with a mean longitudinal velocity of $v_{z}\sim305~\mathrm{ms^{-1}}$~\cite{herburger24a}. This beam intersected the UV laser beam at right angles in the photoexcitation region, which consisted of two parallel circular stainless-steel electrodes separated by a 2.1-cm-wide segmented ring electrode~\cite{scheidegger23a} (see Fig.~\ref{Figure2}(b)). 
This region was surrounded by a double-layer mumetal shield to reduce stray magnetic fields to below 5~mG. 

Doppler-free two-photon spectra of the $(5\mathrm{p})^{5}6\mathrm{p} \, [1/2]_{0}\leftarrow (5\mathrm{p})^{6} \, ^{1}\mathrm{S}_{0}$ transition were recorded at each laser shot by monitoring the Xe$^+$ ions generated by two-photon-resonance-enhanced three-photon ionization [(2+1) REMPI hereafter] as a function of the UV laser frequency. The Xe$^+$ ions were extracted by applying a pulsed electric potential of $255~\mathrm{V}$ to the repeller electrode using a high-potential switch $550~\mathrm{ns}$ after the UV laser pulse. The ions were detected at a microchannel-plate (MCP) detector located at the end of a linear time-of-flight mass spectrometer. The main natural isotopes of Xe and their abundances are $^{128}$Xe, 1.9\%; $^{129}$Xe, 26.4\%; $^{130}$Xe, 4.1\%; $^{131}$Xe, 21.2\%; $^{132}$Xe, 26.9\%; $^{134}$Xe, 10.4\%; $^{136}$Xe, 8.9\%~\cite{mills93a,meija16a}. The spectral resolution of the experiment was sufficient to observe distinct spectral lines for the $^{129}$Xe, $^{131}$Xe, $^{132}$Xe, $^{134}$Xe, and $^{136}$Xe isotopes.

The large pulse-to-pulse fluctuations of the NIR and UV laser outputs (see Fig.~\ref{Figure3}) make it possible to study intensity-dependent effects such as ac-Stark shifts and frequency chirps. To this end, a detection method was implemented which enabled the separate recording, at each laser shot, of the NIR and UV laser intensities and of the integrated signals of the different Xe isotopes. The data was binned according to the amplitude of the UV photodiode signals $V_{\mathrm{ photodiode}}$. In order to fully exploit the dynamical range of the digital oscilloscope used to monitor the ion-time-of-flight spectra, the ion-time-of-flight signal was fed into two separate channels operated at different vertical scales. The first channel, with a strongly magnified vertical scale, was used to analyze weak signals from the low-abundance isotopes and low-intensity UV-laser shots, whereas the second channel, with a compressed vertical scale, was used to analyze strong signals arising from the dominant natural isotopes and high-intensity UV-laser shots. The linearity of the response over the very broad range of signal strengths was tested in separate measurements with intermediate settings.
 
\begin{figure*}[tb!]
\includegraphics[width = \textwidth]{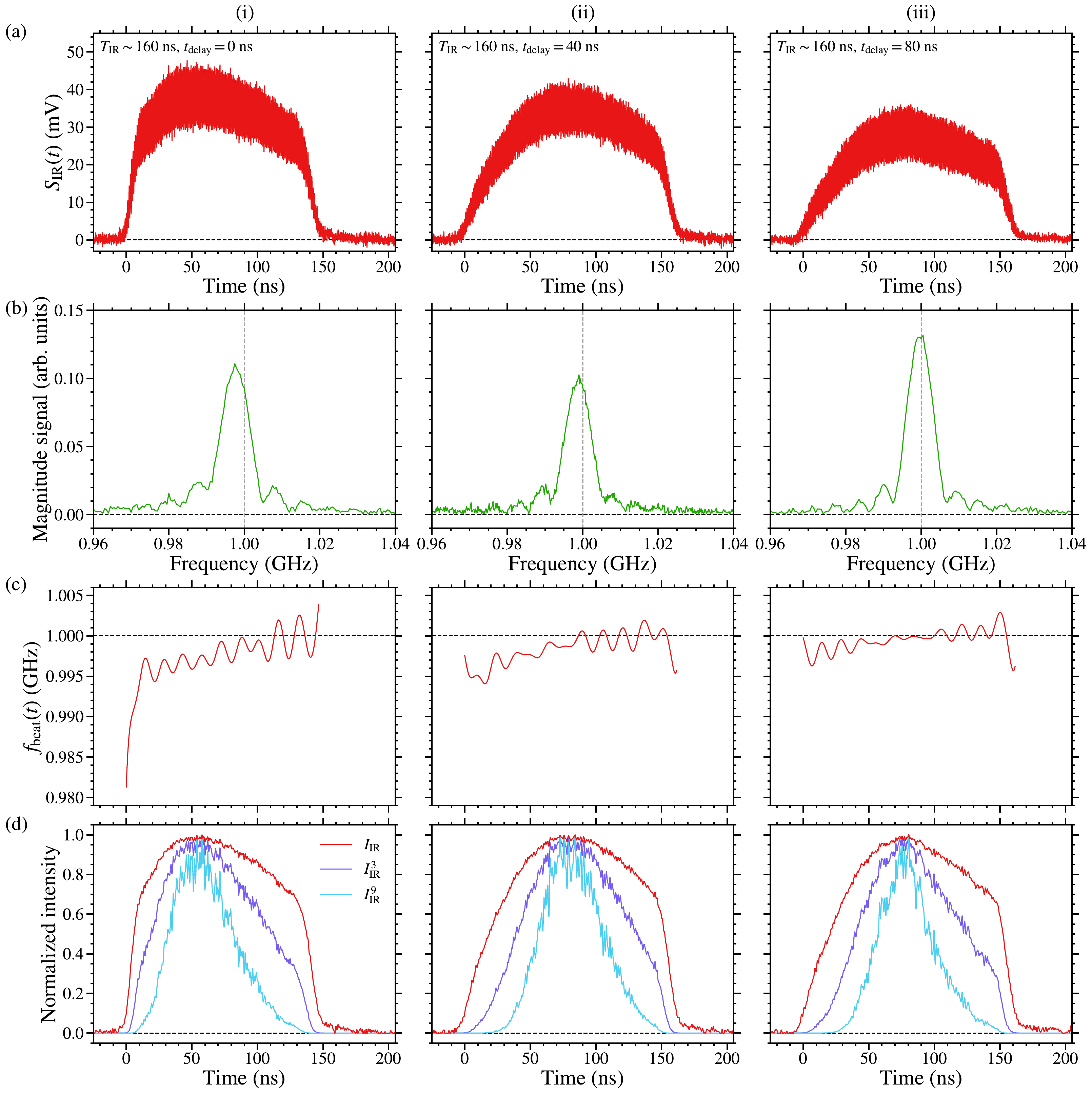} 
\caption{Determination of the frequency chirps of the NIR laser pulses and their dependence on the delay [(i)~$t_{\mathrm{delay}}=0~\mathrm{ns}$, (ii)~$40~\mathrm{ns}$, and (iii)~$80~\mathrm{ns}$] between the Nd:YAG pulse used to pump the Ti:Sa crystals of the bow-tie amplifiers and the seed NIR pulse. (a)~Single-shot beat signal of the amplified NIR pulse against the cw NIR laser output. (b)~Fast-Fourier-transform magnitude spectrum of the beat signal. (c)~Evolution of the instantaneous NIR frequency during the pulse obtained from the fast Fourier transform (see text for details). (d) Envelope of the NIR pulse intensity $I_{\mathrm{IR}}(t)$ (red),  calculated envelopes of the $249~\mathrm{nm}$ UV pulse assuming $I_{\mathrm{UV}}(t) \propto I_{\mathrm{IR}}^{3}(t)$ (purple), and expected unsaturated (2+1) REMPI signal intensity $I_{\mathrm{(2+1)\,REMPI}}(t) \propto I_{\mathrm{IR}}^{9}(t)$ (blue). The intensity scale was normalized to a maximum value of 1 for comparison. \label{Figure4}}
\end{figure*}

If systematic frequency distortions during the pulse amplification were negligible, the frequency of the 
Xe $(5\mathrm{p})^{5}6\mathrm{p} \, [1/2]_{0}\leftarrow (5\mathrm{p})^{6} \, ^{1}\mathrm{S}_{0}$ two-photon transition would be observed at 
\begin{equation}
\label{Eq:Trans_freq_no_chirp}
  2f_{\mathrm{UV}} = 6f_{\mathrm{NIR}} = 6f_{\mathrm{cw}} + 6f_{\mathrm{AOM}}. 
\end{equation}
However, the Nd:YAG-laser pump pulses induce time-dependent changes of the refractive index in the Ti:Sa crystals of the bow-tie amplifiers, which result in frequency shifts and chirps~\cite{seiler05a,liu09b}. To quantify these effects and infer the effective shifts of the two-photon transition frequencies, single-shot measurements of the beat pattern between small fractions of the cw-NIR laser and the pulse-amplified NIR-laser beams were performed using a fast photodiode (Thorlabs DET025A, free-space coupled, nominal bandwidth of $2~\mathrm{GHz}$) coupled to a digital oscilloscope (Lecroy WavePro 760Zi, bandwidth $6~\mathrm{GHz}$). Typical examples of such single-shot beat measurements are presented in Fig.~\ref{Figure4}(a) for pulses of $T_{\mathrm{IR}} \sim 160~\mathrm{ns}$ duration applied after delay times between the Nd:YAG pulse used to pump the Ti:Sa crystals, $t_{\mathrm{delay}}$, of (i)~$0~\mathrm{ns}$, (ii) $40~\mathrm{ns}$, and (iii) $80~\mathrm{ns}$. The measured beat patterns in Fig.~\ref{Figure4}(a) correspond to the combined effects of the envelope of the pulsed laser and the interference between the cw and pulsed IR lasers. In the absence of frequency shifts and chirps, the interference pattern would consist of a regular oscillation at the frequency $f_{\mathrm{beat}} = \left|f_{\mathrm{NIR}} - f_{\mathrm{cw}}\right| = f_{\mathrm{AOM}}$. 
Frequency shifts and chirps during the pulse amplification cause deviations of the instantaneous beat frequency from $f_{\mathrm{AOM}}=1$~GHz and can be reconstructed by Fast Fourier transform (FFT) of the measured photodiode signals. 

Fig.~\ref{Figure4}(b)(i)-(iii) depict the amplitude spectra obtained by FFT of the photodiode signals shown in Fig.~\ref{Figure4}(a)(i)-(iii) around $1~\mathrm{GHz}$. The bandwidth (full width at half maximum) of the laser is about 6~MHz, close to the Fourier-transform limit, and the side bands on each side of the central component are reminiscent of a sinc function and reflect the sharp rising and falling edges of the intensity profile. A red shift of about $2.5~\mathrm{MHz}$ from $f_{\mathrm{AOM}}=1$~GHz is observed when the NIR seed pulse is applied immediately after the Nd:YAG pump pulse [Fig.~\ref{Figure4}(b)(i)]. Panels (b)(ii) and (b)(iii) show that this red shift decreases with increasing time delay between the Nd:YAG pump pulses and the NIR seed pulses, indicating that the refractive index of the Ti:Sa crystal gradually returns close to its original value within about 80~ns, presumably through thermal dissipation of the energy deposited in the Ti.Sa crystals by the Nd:YAG-laser pump pulses.

FFT was also used to reconstruct the IR pulse envelope and the instantaneous beat frequency between the cw and pulsed lasers, following the procedure described in Ref.~\cite{gangopadhyay94a}. The envelopes of the pulse-amplified NIR, $I_{\mathrm{IR}}(t)$, obtained by inverse Fourier transform of the FFT output at frequencies $\leq 0.8~\mathrm{GHz}$ are shown in Fig.~\ref{Figure4}(d)(i)-(iii). The instantaneous beat frequency between the cw and pulsed NIR lasers (see Fig.~\ref{Figure4}(c)) was determined from the phase evolution, $\phi(t)$, using,
\begin{equation}
    f_{\mathrm{beat}}(t) = \frac{1}{2\pi} \frac{\mathrm{d} \phi(t)}{\mathrm{d}t},
\end{equation}
where $\phi(t) = \mathrm{arctan}\left[ \Im(Y(t))/\Re(Y(t)) \right]$ and $Y(t)$ is the inverse Fourier transform of the FFT output filtered around the AOM frequency between $0.925$ and $1.075~\mathrm{GHz}$. In addition to small fluctuations attributed to signal and numerical noise, the evolution of $f_{\mathrm{beat}}(t)$ reveals an initial red shift followed by a gradual return to the unperturbed value of 1~GHz. We also attribute this behavior to the thermal dissipation of the energy deposited in the Ti:Sa crystal by the Nd:YAG laser pulse, and indeed the results depicted in Fig.~\ref{Figure4}(b) and (c) are manifestations of the same phenomenon.
This analysis indicates that chirp-free NIR laser pulses of Fourier-transform-limited bandwidth can be generated provided that the NIR seed pulses are delayed from the Nd:YAG-laser pump pulses by more than about 50 ns. 

\begin{figure}[tb!]
\includegraphics[width = \columnwidth]{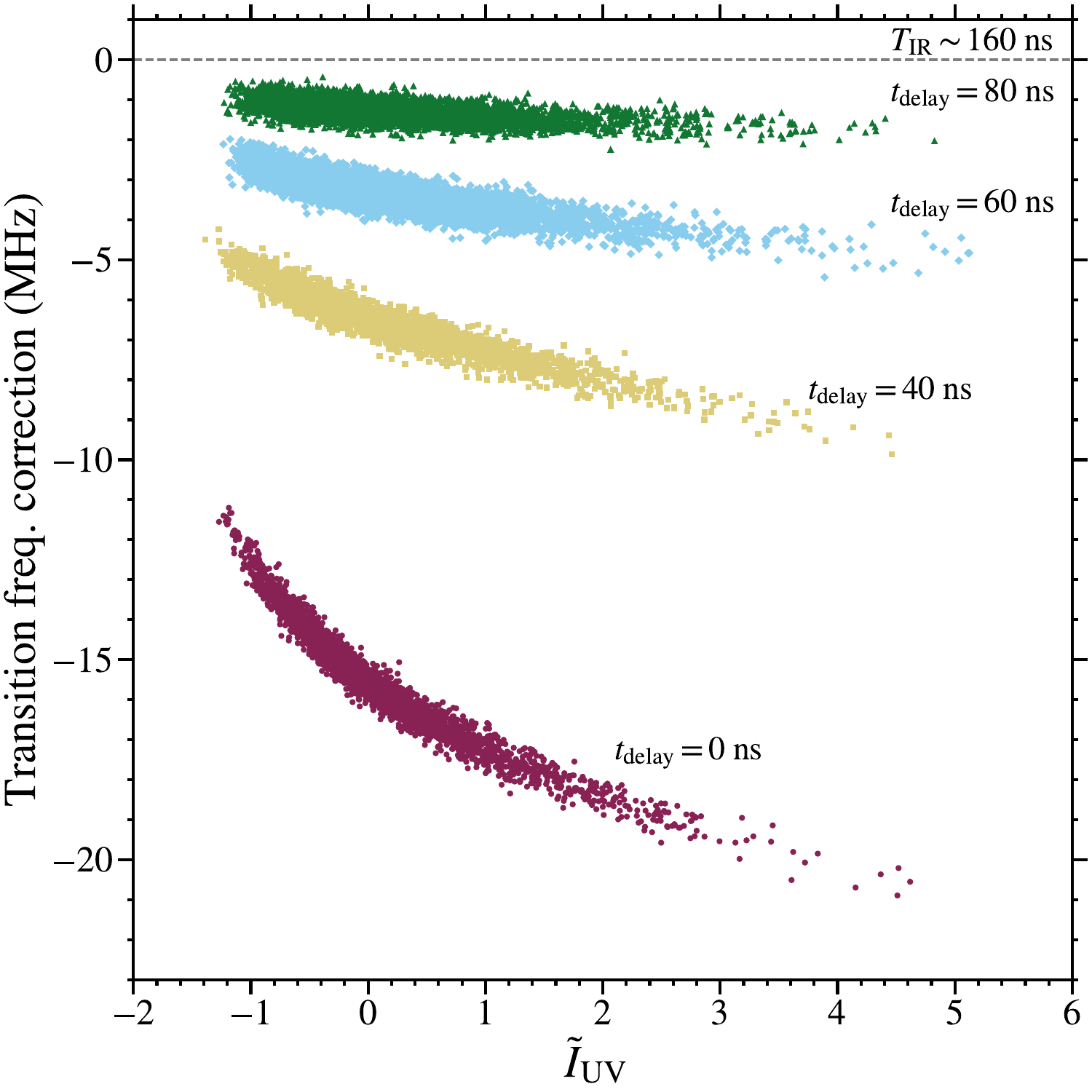} 
\caption{Dependence on the scaled intensity $\tilde{I}_{\mathrm{UV}}$, given by Eq.~(\ref{Eq:I_tilder})  and obtained from the temporal integrated UV intensity distribution $\int I_{\mathrm{UV}}(t) \,\mathrm{d}t$ using $I_{\mathrm{UV}}(t) \propto I_\mathrm{IR}^{3}(t)$ (see Fig.~\ref{Figure4}), of the temporally averaged frequency correction of the $(5\mathrm{p})^{5}6\mathrm{p} \, [1/2]_{0} \leftarrow (5\mathrm{p})^{6} \, ^{1}\mathrm{S}_{0}$ two-photon transition measured by (2+1) REMPI and evaluated with Eq.~(\ref{Eq:Delta_f_trans}) using the instantaneous NIR frequency $f_\mathrm{NIR}(t)$. All data obtained for NIR-pulse durations of $T_{\mathrm{IR}}=$160~ns. \label{Figure5}}
\end{figure}

To assess the effects of the frequency shifts and chirps of the pulsed-amplified NIR laser pulses on the two-photon transition frequency, it needs to be considered that both the frequency upconversion of the NIR pulses to the UV in the BBO crystals and the (2+1) REMPI detection processes are nonlinear processes and thus particularly efficient near the intensity maximum of the NIR pulses. Under conditions where the nonlinear processes are not saturated, the intensity profiles of the UV laser pulses $I_{\mathrm{UV}}(t)$ scale as $I_{\mathrm{IR}}^3(t)$ and the Xe$^+$ signals $S_{\mathrm{Xe^{+}}}(t)$ resulting from the (2+1) REMPI detection scale as $S_{\mathrm{Xe^{+}}}(t) \propto I_{\mathrm{IR}}^9(t)$. The transitions are observed at frequencies corresponding to the sixth harmonic of the NIR pulsed-amplified radiation.

Figure~\ref{Figure4}(d) depicts the experimentally measured envelopes of the amplified NIR laser pulses $I_{\mathrm{IR}}(t)$ normalized to a maximal value of 1 (red) and the normalized envelopes of the functions $I_{\mathrm{UV}}(t)=I_{\mathrm{IR}}^3(t)$ (violet) and $I_{\mathrm{(2+1)\,REMPI}}(t)=I_{\mathrm{IR}}^9(t)$ (blue).
The effective shift of the two-photon UV transition frequency, $\Delta f_{\mathrm{trans}}$, can therefore be determined from the measured NIR beat frequency $f_{\mathrm{beat}}(t)$ and the envelope of the NIR laser pulses $I_{\mathrm{IR}}(t)$ using
\begin{equation}
\label{Eq:Delta_f_trans}
    \Delta f_{\mathrm{trans}} = \left( \frac{6\left(\int f_{\mathrm{beat}}(t) I_{\mathrm{IR}}^{9}(t) \mathrm{d}t \right)}{\int I_{\mathrm{IR} }^{9}(t) \mathrm{d}t}  \right) - 6f_{\mathrm{AOM}},
\end{equation}
i.e., as six times the normalized scalar product of the instantaneous beat frequencies depicted in Fig.~\ref{Figure4}(c) with the corresponding distributions $S_{\mathrm{Xe^{+}}}(t) \propto I_{\mathrm{IR}}^9(t)$ depicted in Fig.~\ref{Figure4}(d). 
The corrected two-photon transition frequency $2f_{\mathrm{UV}}$ thus corresponds to
\begin{equation}
    \label{Eq:Trans_freq_with_chirp}
    2f_{\mathrm{UV}} = 6f_{\mathrm{cw}} + 6f_{\mathrm{AOM}} + \Delta f_{\mathrm{trans}}.
\end{equation}
For any given set of experimental conditions, the fluctuations in the Nd:YAG laser pulse intensities induce shot-to-shot variations of the frequency correction $\Delta f_{\mathrm{trans}}$.
Figure~\ref{Figure5} shows the dependence of the frequency correction $\Delta f_{\mathrm{trans}}$
on the scaled intensity of the pulsed UV-laser radiation, defined by Eq.~(\ref{Eq:I_tilder}) and obtained from the temporally integrated UV intensity distribution $\int I_{\mathrm{UV}}(t) \,\mathrm{d}t$ using $I_{\mathrm{UV}}(t) \propto I_\mathrm{IR}^{3}(t)$ (see Fig.~\ref{Figure4}), for sets of measurements carried out at different delay times $t_{\mathrm{delay}}$ between the Nd:YAG-laser pump pulse and the NIR seed laser pulse. 
The use of the scaled intensity $\tilde{I}_{\mathrm{UV}}$ in this analysis facilitates the comparison of data sets recorded under different experimental conditions, considering only variations of the relative UV pulse intensities. These data show that the frequency shift depends on $t_{\mathrm{delay}}$, as expected from the results presented in Fig.~\ref{Figure4}. They also reveal that
the frequency shift depends on the UV laser intensity. The correlation between the UV intensity and the frequency shift observed in Fig.~\ref{Figure5} originates from the fact that fluctuations in the Nd:YAG pulse energy affect both the refractive index of the Ti:Sa crystals and the efficiency of the pulse amplification and nonlinear frequency upconversion: Stronger Nd:YAG-laser pump pulses do not only cause stronger variations of the refractive index of the crystals, and thus require larger frequency corrections, but they also lead to stronger UV pulses. 

This correlation of frequency shift with UV laser intensity must be considered when analyzing possible ac-Stark shifts of two-photon transitions caused by the UV radiation. The data presented in Fig.~\ref{Figure5} indeed demonstrates the existence of significant UV-intensity-dependent frequency shifts that are not caused by the ac-Stark effect but by two distinct (noncausal) correlated effects in Ti:Sa pulsed amplification systems. Such shifts are not restricted to Ti:Sa pulse-amplification schemes but are an intrinsic feature of combined pulse-amplification and nonlinear optical schemes. 

\section{Results}
\label{Sec:Results}

\begin{figure}[tb!]
\includegraphics[width = \columnwidth]{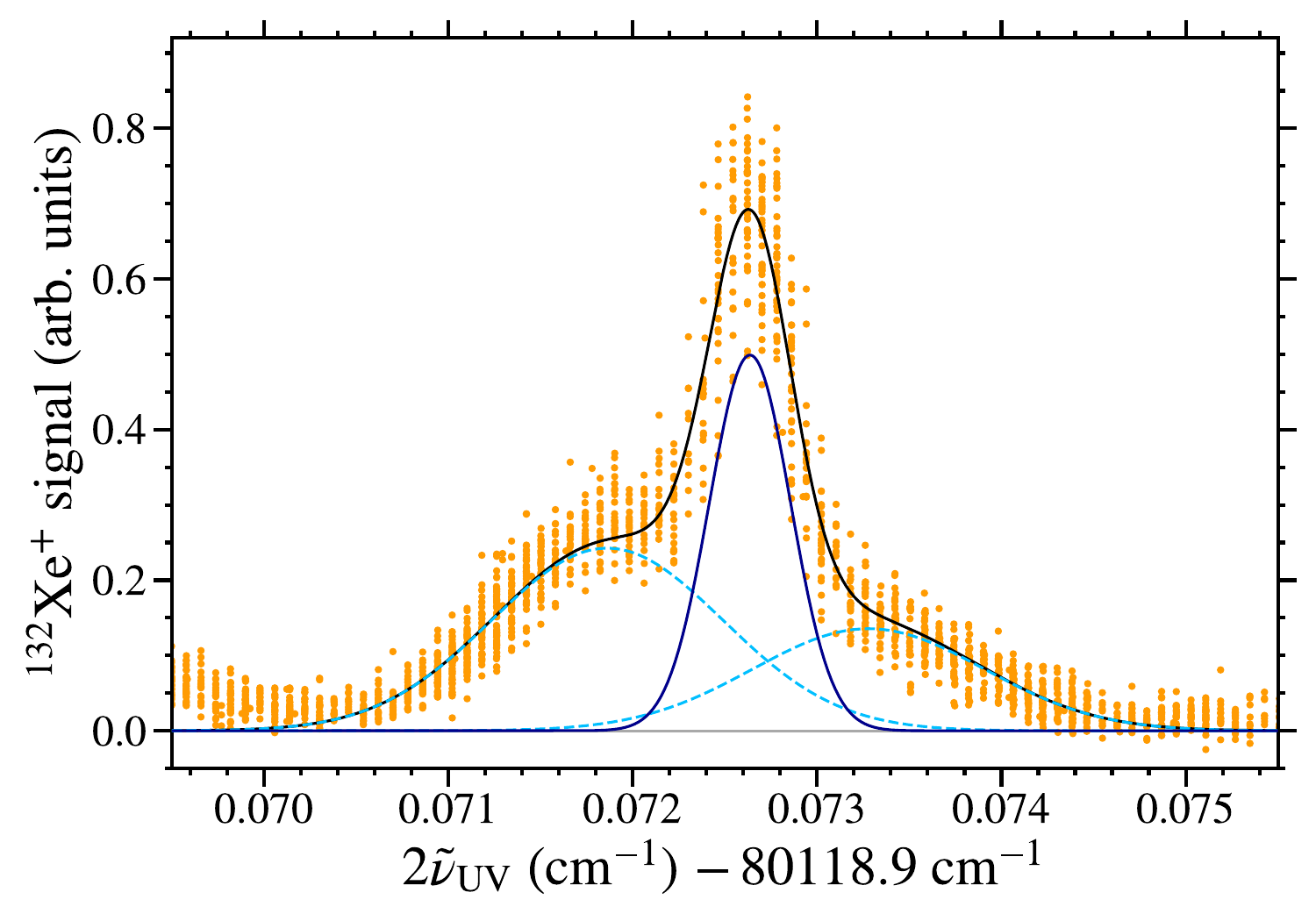}
\caption{Line shape of the $(5\mathrm{p})^{5}6\mathrm{p} \, [1/2]_{0}\leftarrow (5\mathrm{p})^{6} \, ^{1}\mathrm{S}_{0}$ two-photon transition of $^{132}$Xe recorded by (2+1) REMPI spectroscopy with counterpropagating UV laser beams. The orange dots and the full black line represent the experimental data and the fitted line profile (for only the $^{132}$Xe isotope), respectively. The dashed light-blue lines represent the contributions to the intensity distribution resulting from the absorption of two photons from the forward-propagating (low-frequency component) and retroreflected (high-frequency component) beams. The dark-blue line corresponds to the Doppler-free contribution with one photon absorbed from each of the counterpropagating UV beams. Note that the wavenumbers on the horizontal axis do not include correction for the frequency shifts given by Eq.~\ref{Eq:Delta_f_trans}. \label{Figure6}}
\end{figure}

The counterpropagating arrangement of the UV laser beams used to record spectra of the $(5\mathrm{p})^{5}6\mathrm{p} \, [1/2]_{0}\leftarrow (5\mathrm{p})^{6} \, ^{1}\mathrm{S}_{0}$ 
two-photon resonance of Xe leads to characteristic line profiles, illustrated in Fig.~\ref{Figure6}. These profiles consist of a narrow 
Doppler-free Gaussian component, corresponding to the absorption of one photon from each of the forward-propagating and retroreflected UV laser beams, and two Doppler-broadened Gaussian components resulting from the absorption of two photons from either the forward-propagating beam or the retroreflected beam. Because of a small deviation from 90$^\circ$ of the angle between the Xe and UV-laser beams, the Doppler-broadened components appear as two distinct Doppler-shifted lines with opposite Doppler shift beside the Doppler-free component. Intensity losses at the mirror and the exit window of the vacuum chamber significantly attenuate the Doppler-broadened line of the retroreflected UV beam. 

The overall line profile can be accurately reproduced by fitting three Gaussian line shapes to the measured intensity distribution, one for the central Doppler-free component and two for the side bands, as illustrated by the solid black line in Fig.~\ref{Figure6}.
The full width at half maximum of the central Doppler-free component ($\varGamma_{(2+1)\, {\mathrm{REMPI}}} \sim 15~{\mathrm{MHz}}$, solid dark-blue line) primarily reflects the upconversion of the bandwidth $\varGamma_{\mathrm{NIR}}\sim 6~\mathrm{MHz}$ of the NIR radiation to $\varGamma_{(2+1)\, {\mathrm{REMPI}}}=\sqrt{6}\varGamma_{\mathrm{NIR}}\sim14.7$~MHz. The full width at half maximum of the Doppler-broadened side bands ($\varGamma_{\mathrm{Doppler}}\sim45~\mathrm{MHz}$, dashed light-blue lines) reflects the distribution of transverse velocities of the Xe atoms in the supersonic beam. 

\begin{figure}[tb!]
\includegraphics[width =\columnwidth]{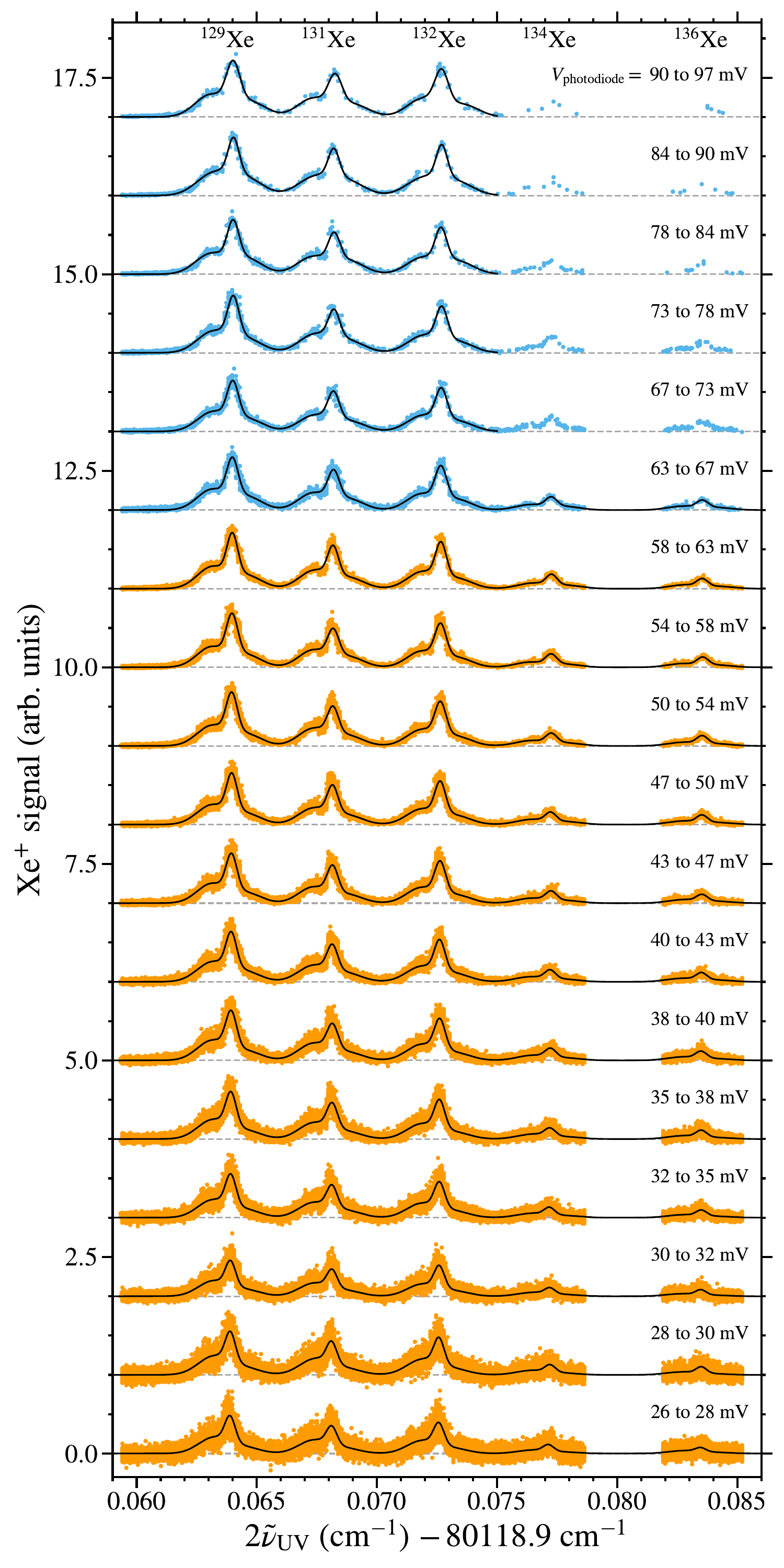}
\caption{Spectra of the $(5\mathrm{p})^{5}6\mathrm{p} \, [1/2]_{0} \leftarrow (5\mathrm{p})^{6} \, ^{1}\mathrm{S}_{0}$ two-photon transition of the main isotopes of Xe, as labeled at the top of the figure, recorded using $T_{\mathrm{IR}}=240~\mathrm{ns}$ and $t_{\mathrm{delay}}=0~\mathrm{ns}$. The wavenumbers on the horizontal axis do not include the correction for the frequency shifts given in Eq.~\ref{Eq:Delta_f_trans}. The spectra are ordered along the vertical axis according to the signal $V_{\mathrm{photodiode}}$ of the photodiode used to monitor the relative UV-laser intensities, as indicated on the right-hand side. The colored dots and full lines correspond to experimental data and fitted line profiles, respectively. The spectra displayed in orange and blue were obtained by using Xe$^+$ ion signals monitored using the oscilloscope channels with magnified and reduced vertical scales, respectively. See text for details.
\label{Figure7}}.
\end{figure}

\begin{figure}[tb!]
\includegraphics[width = \columnwidth]{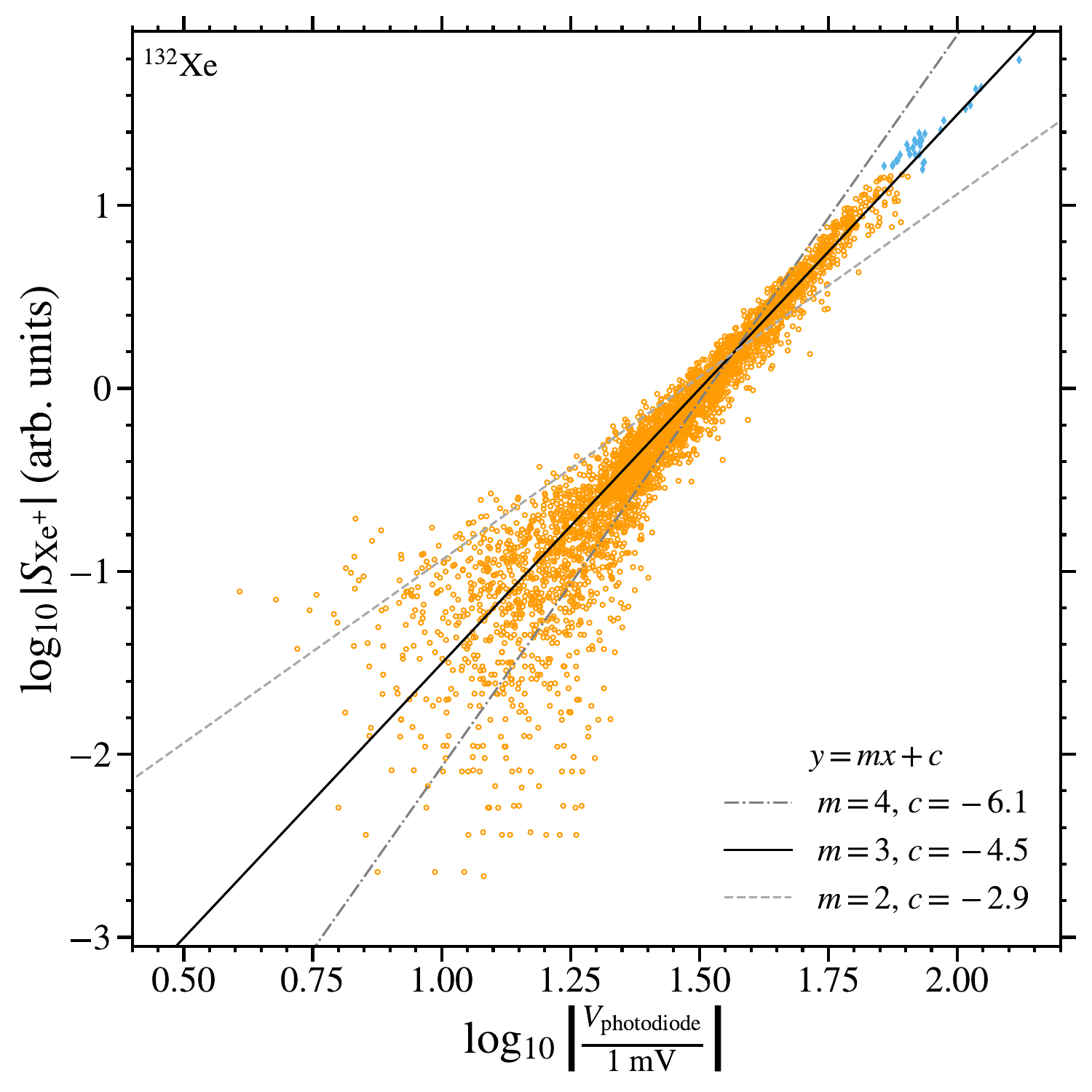}
\caption{Log-log plot of the (2+1) REMPI $^{132}$Xe$^{+}$ ion signal $S_{\mathrm{Xe^{+}}}$ near its maximum as a function of 
the signal $V_{\mathrm{photodiode}}$ of the photodiode used to monitor the relative UV-laser intensities. The orange and blue dots represent the $^{132}$Xe$^+$ ion signal monitored using the oscilloscope channels with magnified and reduced vertical scales, respectively.
The dash-dotted, full, and dashed lines correspond to fits of linear functions with slopes $m$ of 2, 3, and 4, respectively. \label{Figure8}}
\end{figure}

\begin{figure}[tb!]
\includegraphics[width = \columnwidth]{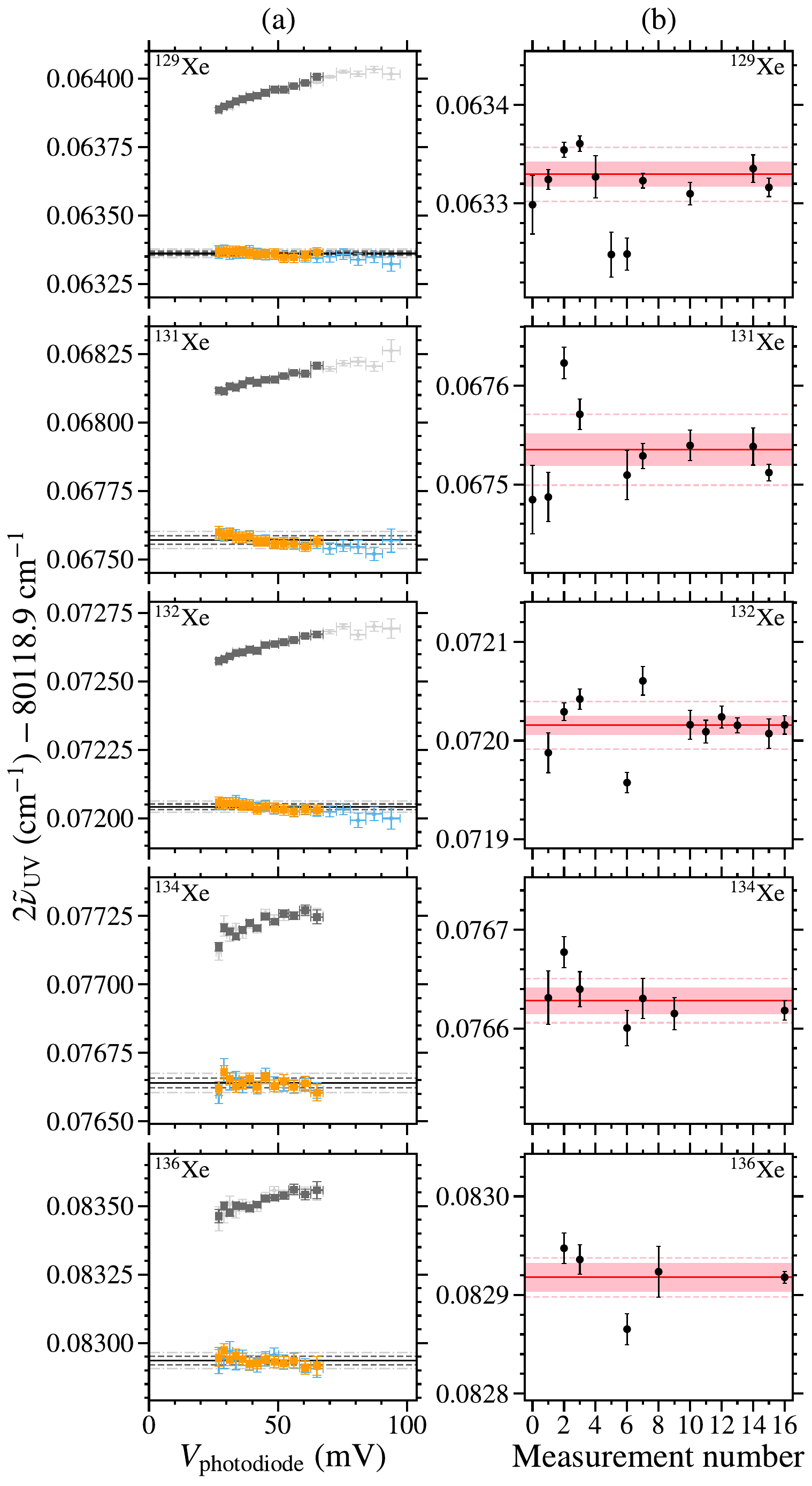}
\caption{(a) Comparison of the central frequencies of the $(5\mathrm{p})^{5}6\mathrm{p} \, [1/2]_{0} \leftarrow (5\mathrm{p})^{6} \, ^{1}\mathrm{S}_{0}$ two-photon transition of the main isotopes of Xe without (gray dots) and after (colored dots) correcting for the frequency shifts $\Delta f_{\mathrm{trans}}$ originating from the intensity-dependent frequency chirps using Eq.~\ref{Eq:Delta_f_trans}. The experimental data for this example were obtained for $T_{\mathrm{IR}} = 240~\mathrm{ns}$ and $t_{\mathrm{delay}}= 0~\mathrm{ns}$ and are shown in Fig.~\ref{Figure7}. The full, dashed, and dash-dotted horizontal lines represent the weighted mean of the corrected central frequencies, the $1\sigma$, and $2\sigma$ confidence intervals, respectively. (b) Overview of the results of all measurements of the two-photon transition frequencies carried out for different combinations of $T_{\mathrm{IR}}$ and $t_{\mathrm{delay}}$ values. The weighted means, the square root of the weighted variance (equivalent to $1\sigma$), and the standard error of the weighted mean are indicated by the solid red lines, the pink dashed lines, and red shaded areas, respectively. \label{Figure9}}
\end{figure}

Spectra of the $(5\mathrm{p})^{5}6\mathrm{p} \, [1/2]_{0}\leftarrow (5\mathrm{p})^{6} \, ^{1}\mathrm{S}_{0}$ two-photon transition were measured for the most abundant Xe isotopes, i.e., $^{129}$Xe, $^{131}$Xe, $^{132}$Xe, $^{134}$Xe, and $^{136}$Xe. Because of the large UV-laser intensity fluctuations (see Fig.~\ref{Figure3}), the ion signals were monitored on a shot-to-shot basis and sorted according to the integrated UV-laser photodiode signal $V_{\mathrm{photodiode}}$.
As illustration, Fig.~\ref{Figure7} shows the (2+1) REMPI spectra obtained after binning the Xe$^+$ signal ($S_{\mathrm{Xe}^{+}}$) into 18 categories with $V_{\mathrm{photodiode}}$ ranging from 26~mV to 97~mV, as indicated on the right of the figure. These spectra all have the same structure and reveal five dominant spectral lines, one for each of the dominant Xe isotopes, that have the characteristic three-component line shape discussed above (see Fig.~\ref{Figure6}). The dependence of the Xe$^+$ signal strength on the UV-laser intensity is shown in a log-log plot in Fig.~\ref{Figure8} for $^{132}$Xe and demonstrates the $S_{\mathrm{Xe}^{+}} \propto I_{\mathrm{UV}}^{3}$ relation expected for an unsaturated (2+1) REMPI excitation, as discussed in Sec.~\ref{Sec:Experiment}.

Detailed inspection of the spectra presented in Fig.~\ref{Figure7} reveals systematic shifts of the line positions with increasing UV-laser intensities.
To quantify these intensity-dependent shifts, the spectra obtained from the different bins (see Fig.~\ref{Figure7}) were analyzed separately.
For each spectrum, five three-component line profiles of the type described above (see Fig.~\ref{Figure6}), i.e., one for each of the dominant isotopes, were fitted to determine the line centers of the Doppler-free components. 
In the fitting procedure, the values of $\varGamma_{\mathrm{Doppler}}$, $\varGamma_{(2+1)\, {\mathrm{REMPI}}}$, the energy separation between the two Doppler-broadened side bands and their relative intensities were assumed to be the same for all isotopes. 
The results of these fits are represented by the solid black lines in Fig.~\ref{Figure7} and perfectly reproduce the experimentally measured intensity distributions. 

The center frequencies of the Doppler-free components of each isotope determined in these fits are plotted against $V_{\mathrm{photodiode}}$ as gray data points in Fig.~\ref{Figure9}(a). The values of $V_{\mathrm{photodiode}}$ correspond to the middle positions of the respective UV-laser-intensity bin ranges and the horizontal error bars represent their widths. The data points in pale gray correspond to the data obtained with the reduced oscilloscope scale, which were only reliable at the largest $V_{\mathrm{photodiode}}$ ranges for the three most abundant isotopes for this example. The spectra presented in Fig.~\ref{Figure9}(a) reveal a shift of the two-photon wavenumbers with increasing UV-laser intensity, 
the differences between the spectra recorded at the lowest and highest UV-laser intensities being about $1.5\times 10^{-5}$~cm$^{-1}$ (5~MHz). 

To assess whether the transition frequencies were affected by ac-Stark shifts, it was necessary to first correct for the intensity-dependent laser-frequency shifts $\Delta f_{\mathrm{trans}}$ caused by the shot-to-shot intensity fluctuations (see Eq.~(\ref{Eq:Delta_f_trans}) and Fig.~\ref{Figure5}). Because we only measured the relative UV-laser intensities and the Xe$^+$ signals when recording spectra such as those displayed in Fig.~\ref{Figure7}, the effects of the temporal evolution of the frequencies during the individual laser shots had to be reconstructed {\it a posteriori} using the calibration of $\Delta f_{\mathrm{trans}}$ carried out in the analysis of frequency chirps and shifts (see Figs.~\ref{Figure4} and~\ref{Figure5}). The correction procedure is based on the use of the scaled intensities $\tilde{I}_{\mathrm{UV}}$ (see Eq.~(\ref{Eq:I_tilder})) and was performed in two steps. In the first step, the intensity distributions of the UV-laser shots monitored at the UV photodiode when recording the spectra were used to determine the corresponding $\tilde{I}_{\mathrm{UV}}$ parameters. These parameters, which were also determined in the analysis of frequency chirps and shifts using $I_{\mathrm{UV}}=\int I_{\mathrm{UV}}(t) \,\mathrm{d}t
\propto \int  I_\mathrm{IR}^{3}(t) \,\mathrm{d}t$,
were then used, in the second step, to extract the $\Delta f_{\mathrm{trans}}$ values and their uncertainties (see Fig.~\ref{Figure5}) from a statistical analysis of the data corresponding to each $V_{\mathrm{photodiode}}$ range.
The colored data points in Fig.~\ref{Figure9}(a) depict the corrected two-photon transition frequencies determined for each isotope and each $V_{\mathrm{photodiode}}$ range in Fig.~\ref{Figure7}. These corrected frequencies do not reveal significant residual intensity-dependent shifts within the statistical limits of the measurements ($\sim 6\times 10^{-6}$~cm$^{-1}$ or $\sim 200$~kHz), which in turn implies that ac-Stark shifts are negligible in the range of UV-laser intensities used to record the spectra shown in Fig.~\ref{Figure7}.
The transition frequencies and their statistical uncertainties were therefore determined by calculating the weighted mean and the square root of the weighted variance (equivalent to $1\sigma$) of the corrected Doppler-free transition frequencies, respectively. 

Multiple measurements of the $(5\mathrm{p})^{5}6\mathrm{p} \, [1/2]_{0}\leftarrow (5\mathrm{p})^{6} \, ^{1}\mathrm{S}_{0}$ transition frequency were performed for different combinations of $T_{\mathrm{IR}}$ and $t_{\mathrm{delay}}$, which involved up to $300000$ experimental cycles each. These measurements were analyzed using the procedure described above, and the results of all measurements are shown as black data points in Fig.~\ref{Figure9}(b). The final values of the transition frequencies and their statistical uncertainties were determined as the weighted means (solid red lines in Fig.~\ref{Figure9}(b)) and standard errors of the weighted means (red-shaded areas). 

\begin{table}[t!]
    \caption{\label{Table0}\small Overview of systematic shifts and uncertainties. All values are in kHz.}
    \begin{ruledtabular}
    \begin{tabular}{lcc}
        Source & \multicolumn{1}{c}{\textrm{Shift}} & \multicolumn{1}{c}{\textrm{Uncertainty}} \\
        \colrule
        Residual first-order Doppler shift\footref{Table1_footnote}  &                                                                 & $660$ \\
         Frequency Calibration               &                                                                 & $40$ \\
        Second-order Doppler shift\footnote{Exact value depends on the transition frequency.\label{Table1_footnote}}          &  $-1.2$ & \\
        Photon-recoil shift                 & $\ll 1$                                          & \\
        ac-Stark shift\footnote{Neglected in the final systematic uncertainty because of statistical averaging, see text.}                     &                                                                 &  $< 100$ \\
        dc-Stark shift                      &                                                                 &  $\ll 1$ \\
        Zeeman shift                        &                                                                 &  $\ll 1$ \\
        \colrule
        $\sigma_\mathrm{sys}$               &                                                                 & $700$ \\
    \end{tabular}
    \end{ruledtabular}
\end{table}

Table~\ref{Table0} summarizes the systematic errors and uncertainties considered in the analysis. Systematic shifts of the transition frequencies from Zeeman or Stark shifts are expected to be negligible under the experimental conditions used for the measurements, considering that stray electric and magnetic fields in the photoexcitation region do not exceed 5~mV/cm and 5 mG, respectively. The residual photon-recoil shift $\left[ \frac{2hf^{2}_{\mathrm{UV}}}{mc^{2}}\sin^{2} \left(\frac{\theta_{\mathrm{max}}}{2} \right) \right]$ and the second-order Doppler shift [$\Delta(2 f_{\mathrm{UV}}) = - (2f_{\mathrm{UV}}) v_z^2/(2c^2)$] do not contribute significantly to the uncertainty in these measurements. The residual first-order Doppler shift arising from the maximum possible deviation angle $\theta_{\mathrm{max}}$ from exact $180^{\circ}$ retroreflection introduces a systematic uncertainty of
\begin{equation}
   \Delta(2f_{\mathrm{UV}}) = \left(\frac{v_{z}}{c} \right) f_{\mathrm{UV}} \sin{\theta_{\mathrm{max}}}.
\end{equation}
to the total transition frequency uncertainty. 
For the experiments reported here, $\Delta(2f_{\mathrm{UV}}) = 0.66~\mathrm{MHz}$ ($\equiv 2.2 \times10^{-5}~\mathrm{cm^{-1}}$) and represents the primary source of systematic uncertainty. The stability of the lock of the cw NIR frequency to the frequency comb is estimated to contribute $\sim0.04~\mathrm{MHz}$ to the total $2f_{\mathrm{UV}}$ systematic uncertainty.  The final values of the $(5\mathrm{p})^{5}6\mathrm{p} \, [1/2]_{0}\leftarrow (5\mathrm{p})^{6} \, ^{1}\mathrm{S}_{0}$ transition frequencies and the corresponding total uncertainties are given in Table~\ref{Table1} for the five main natural isotopes of Xe. The measurement method thus enables the determination of two-photon transition frequencies with sub-MHz accuracy.

\begin{table}[tb!]
\caption{Wavenumbers and frequencies of the $(5\mathrm{p})^{5}6\mathrm{p} \, [1/2]_{0} \leftarrow (5\mathrm{p})^{6} \, ^{1}\mathrm{S}_{0}$ two-photon transitions of the most abundant Xe isotopes determined in the present work. The numbers in parentheses represent the experimental uncertainties, which include the contributions from the systematic uncertainties (see text for details).  \label{Table1}}
\begin{ruledtabular}
\begin{tabular}{l D{.}{.}{5.10} D{.}{.}{10.6}}
Isotope  & \multicolumn{1}{c}{\textrm{$2\tilde{\nu}_{\mathrm{UV}}$} ($\mathrm{cm^{-1}}$)} & \multicolumn{1}{c}{\textrm{$2f_{\mathrm{UV}}$ ($\mathrm{MHz}$)}}    \\
\colrule \vspace{-0.7em} \\
$^{129}$Xe & 80118.963330(26) & 2401906094.90(79) \\
$^{131}$Xe & 80118.967535(28) & 2401906220.98(85) \\
$^{132}$Xe & 80118.972016(25) & 2401906355.30(75) \\
$^{134}$Xe & 80118.976628(27) & 2401906493.58(80) \\
$^{136}$Xe & 80118.982918(27) & 2401906682.14(82) \\
\end{tabular}
\end{ruledtabular}
\end{table}

\begin{figure}[tb!]
\includegraphics[width = \columnwidth]{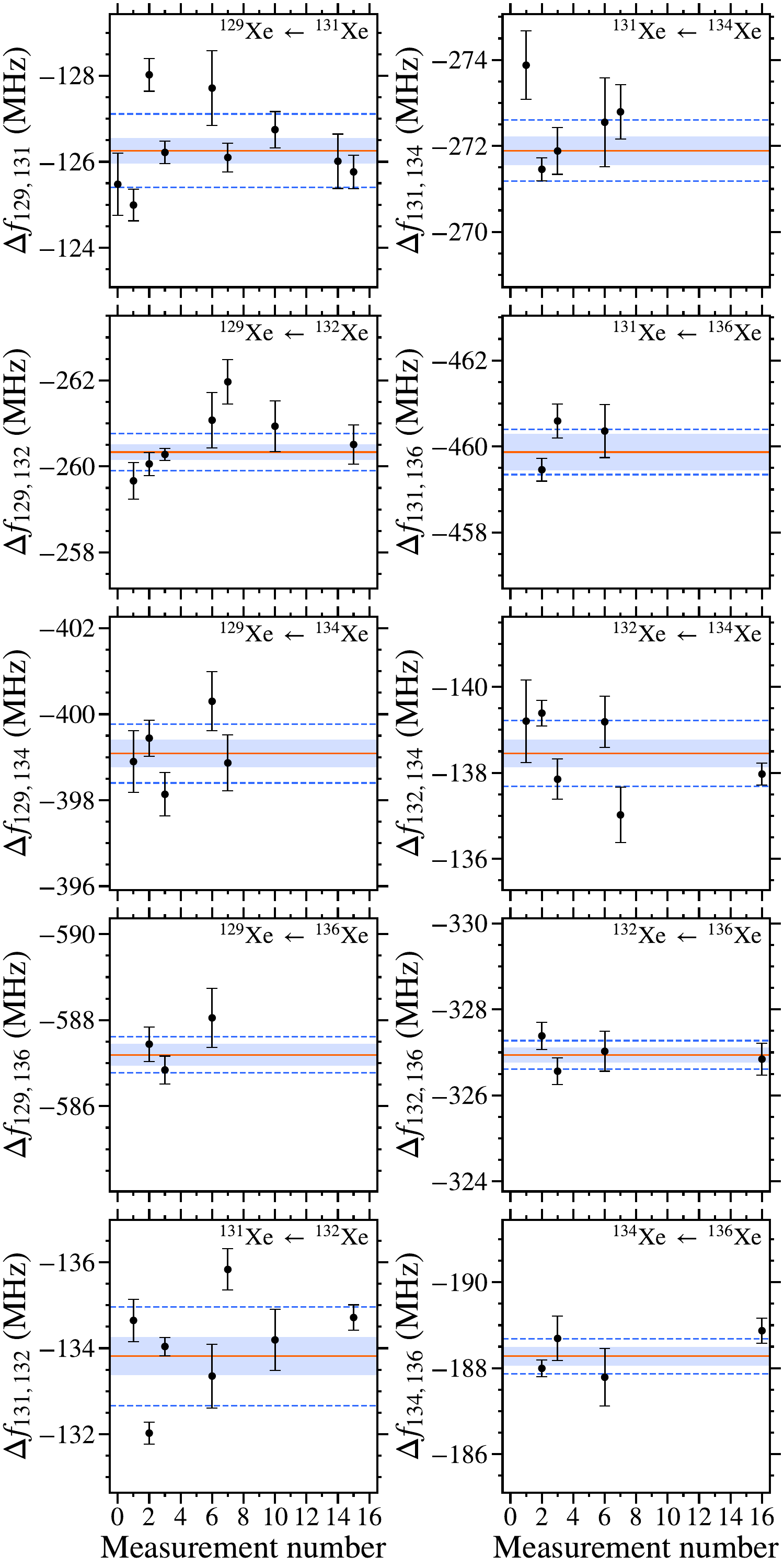}
\caption{Isotopic shifts $\Delta f_\mathrm{A,B} = 2f_{\mathrm{A}} - 2f_{\mathrm{B}}$  of the $(5\mathrm{p})^{5}6\mathrm{p} \, [1/2]_{0} \leftarrow (5\mathrm{p})^{6} \, ^{1}\mathrm{S}_{0}$ two-photon transition of Xe obtained for the isotope pairs (A,B) for which accurate transition frequencies could be determined experimentally. The solid orange lines, the blue dashed lines and the blue shaded regions designate the weighted means, the square root of the weighted variance (equivalent to $1\sigma$), and the error of the weighted means, respectively.
\label{Figure10}}
\end{figure}

The systematic uncertainties of the two-photon transition frequencies cancel out when determining isotopic shifts $\Delta f_\mathrm{A,B} = 2f_{\mathrm{A}} - 2f_{\mathrm{B}}$. Moreover, the intensity-dependent frequency shifts for a given measurement do not need to be corrected for because the corrections for each range of UV laser photodiode signal $V_{\mathrm{photodiode}}$ are the same for all isotopes. Consequently, the uncertainties of isotopic shifts are solely of statistical origin.
For any particular measurement, the isotopic shifts and their uncertainties were derived as the differences between the uncorrected Doppler-free frequencies of a pair of isotopes and their square-rooted weighted variance (equivalent to $1\sigma$) considering spectra recorded in all $V_{\mathrm{photodiode}}$ ranges. 
The results of this analysis for all measurements of the $(5\mathrm{p})^{5}6\mathrm{p} \, [1/2]_{0} \leftarrow (5\mathrm{p})^{6} \, ^{1}\mathrm{S}_{0}$ two-photon transition frequencies of Xe carried out in the present investigation are presented as black dots with error bars in Fig.~\ref{Figure10}, where the orange horizontal lines represent the weighted mean values, the dashed blue lines the standard deviations, and the blue shaded areas the standard errors of the weighted means. Because of the cancellation of systematic uncertainties, the isotopic shifts could be determined with an accuracy of $\sim$200~kHz, limited by the statistical uncertainties.
The final values of the isotopic shifts $\Delta f_\mathrm{A,B}$ are listed in Table~\ref{Table2}.

\begin{table}[tb!]
\caption{Isotopic shifts $\Delta f_\mathrm{A,B} = 2f_{\mathrm{A}} - 2f_{\mathrm{B}}$ of the $(5\mathrm{p})^{5}6\mathrm{p} \, [1/2]_{0} \leftarrow (5\mathrm{p})^{6} \, ^{1}\mathrm{S}_{0}$ two-photon transition frequencies of Xe obtained for the isotope pairs (A,B) and corresponding to the data presented in Fig.~\ref{Figure10}.  \label{Table2}}
\begin{ruledtabular}
\begin{tabular}{lc D{.}{.}{4.6}}
Isotope A  & Isotope B & \multicolumn{1}{c}{\textrm{$\Delta f_{\mathrm{A,B}}$ ($\mathrm{MHz}$)}}    \\
\colrule \vspace{-0.7em} \\
$^{129}$Xe & $^{131}$Xe & -126.26(29) \\
$^{129}$Xe & $^{132}$Xe & -260.33(17) \\
$^{129}$Xe & $^{134}$Xe & -399.09(31) \\
$^{129}$Xe & $^{136}$Xe & -587.19(25) \\
$^{131}$Xe & $^{132}$Xe & -133.81(44) \\
$^{131}$Xe & $^{134}$Xe & -271.89(32) \\
$^{131}$Xe & $^{136}$Xe & -459.87(41) \\
$^{132}$Xe & $^{134}$Xe & -138.45(32) \\
$^{132}$Xe & $^{136}$Xe & -326.94(17) \\
$^{134}$Xe & $^{136}$Xe & -188.28(21) \\
\end{tabular}
\end{ruledtabular}
\end{table}

\section{Discussion}
\label{Sec:Discussion}

The measurements of the $(5\mathrm{p})^{5}6\mathrm{p} \, [1/2]_{0}\leftarrow (5\mathrm{p})^{6} \, ^{1}\mathrm{S}_{0}$ transition frequencies in Xe presented in Section~\ref{Sec:Results} contribute to the clarification of the origin of the {\mbox{$\sim1~\mathrm{GHz}$}} discrepancy between a recent precision measurement of the ionization energy of Xe determined from the metastable ($5$p)${^5}$ $6$s $[3/2]_{2}$ state  by Herburger \emph{et al.}~\cite{herburger24a} and the first-ionization energy of the ground state~of Xe recommended in the NIST atomic data base (isotope-abundance-weighted value of 97833.787(11)~cm$^{-1}$)~\cite{nist_asd_template}. The current NIST values~\cite{nist_asd_template} relies on the combination of (i)~the frequency of the $(5\mathrm{p})^{5}(8\mathrm{d})[1/2]_{1}\leftarrow (5\mathrm{p})^{6} \, ^{1}\mathrm{S}_{0}$ transition in the vacuum-ultraviolet range~\cite{brandi01a}, (ii)~the ionization energy of the metastable ($5$p)${^5}$ $6$s$^\prime$ $[1/2]_{0}$ state reported by Knight {\it et al.} from an extrapolation of Rydberg series~\cite{knight85a}, and (iii)~IR transitions frequencies from Ref.~\cite{humphreys70a} used to connect the $(5\mathrm{p})^{5}(8\mathrm{d})[1/2]_{1}$ and ($5$p)${^5}$ $6$s$^\prime$ $[1/2]_{0}$ states (see Ref.~\cite{brandi01a} for details). 

\begin{figure}[tb!]
\includegraphics[width = \columnwidth]{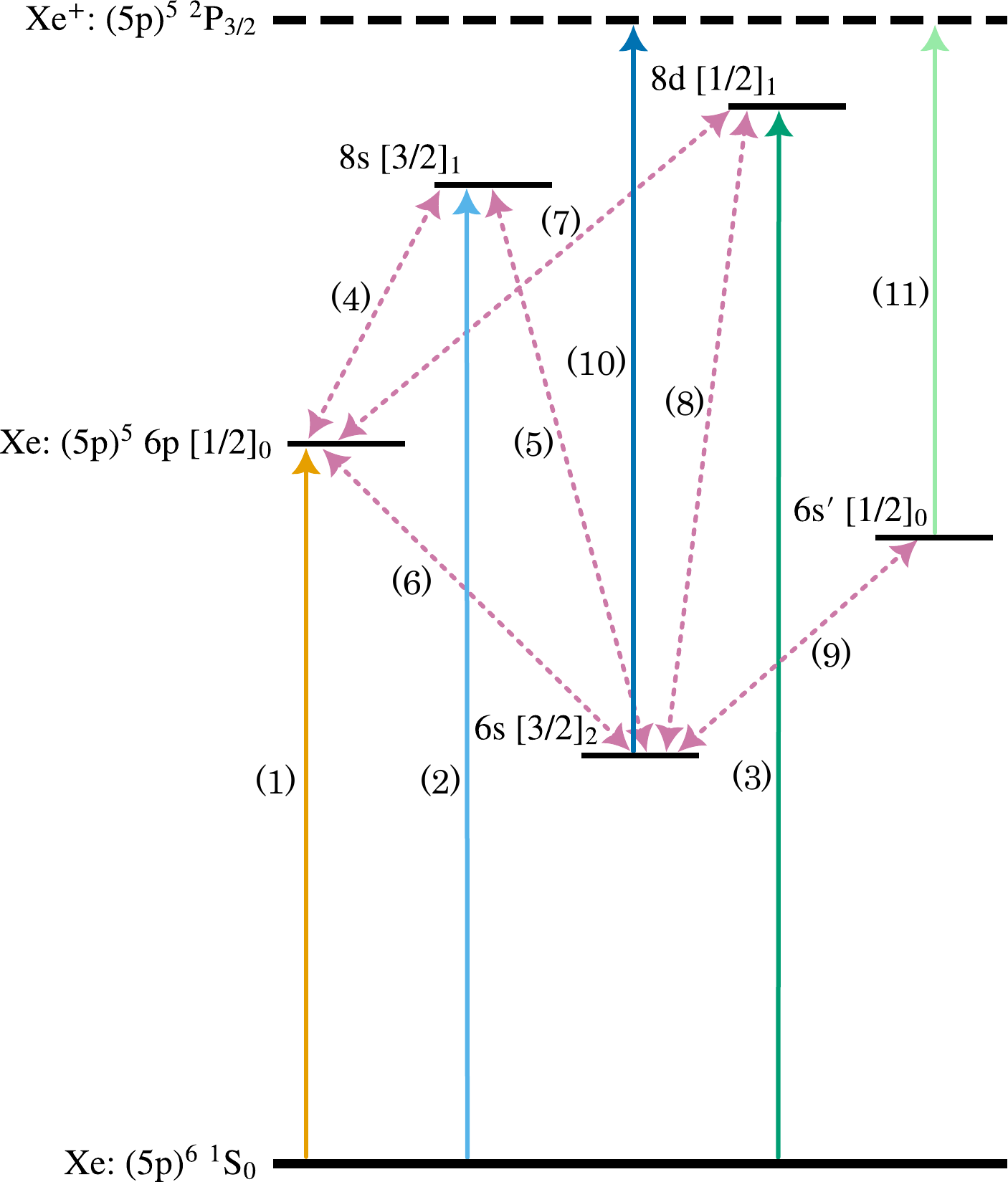}
\caption{Level diagram illustrating the energy intervals used to determine the ${\mathrm{^{136}Xe}}^+\ (5\mathrm{p})^{5}\ ^2{\mathrm{P}}_{3/2}\leftarrow {\mathrm{^{136}Xe}}\ (5\mathrm{p})^{6} \, ^{1}\mathrm{S}_{0}$ ionization energy of xenon from a redundant set of energy intervals connecting the $(5\mathrm{p})^{6} \, ^{1}\mathrm{S}_{0}$ ground state to the $6{\mathrm{p}}[1/2]_0$ (orange, this work), $8{\mathrm{s}}[3/2]_1$ (light blue, \cite{brandi01a}), and $8{\mathrm{d}}[1/2]_1$ (dark green, \cite{dreissen19a,dreissen20a}) low-lying Rydberg states, and connecting the $6{\mathrm{s}}[3/2]_2$ (dark blue, \cite{herburger24a}) and $6{\mathrm{s}}^\prime[1/2]_0$ (pale green, \cite{knight85a}) low-lying Rydberg states to the ${\mathrm{Xe}}^{+}\ ^{2}{\mathrm{P}}_{3/2}$ ionization threshold. Redundancy in the network is provided by intervals (pink dashed lines) obtained from different combinations of NIR and visible transition frequencies \cite{humphreys70a,littlefield74a},
as detailed in the supplemental material~\cite{SupplementalMaterial}. Numbers correspond to those used in Table~\ref{Table3} to label the energy intervals.
\label{Figure11}}
\end{figure}

The energy-level diagram illustrating the energy intervals $\Delta E_{(i)}$ used in the reevaluation of the ionization energy of Xe in the present work is presented in Fig.~\ref{Figure11}. The values of the corresponding wavenumbers $\Delta E_{(i)}/(hc)$ are listed in Table~\ref{Table3}. Three intervals [Intervals (1), (2) and (3)], indicated by orange, light-blue and dark green arrows, connect the $(5\mathrm{p})^{6}\, ^{1}\mathrm{S}_{0}$ ground state to low-lying electronic states, the $(5\mathrm{p}){^5} \, 6\mathrm{p} \, [1/2]_{0}$~[this work], $(5\mathrm{p}){^5}(8\mathrm{d})[1/2]_{1}$~\cite{brandi01a}, and $(5\mathrm{p}){^5}(8\mathrm{s})[3/2]_{1}$~\cite{dreissen19a,dreissen20a} states, respectively. 
Two intervals [Intervals (10) and (11)], indicated by dark-blue and light-green arrows, connect the first ($(5\mathrm{p})^{5}\ ^2{\mathrm{P}}_{3/2}$) ionization threshold of Xe to low-lying electronic states of Xe, the $(5\mathrm{p}){^5}(6\mathrm{s})[3/2]_{2}$~\cite{herburger24a} and $(5\mathrm{p}){^5}(6\mathrm{s}^\prime)[1/2]_{0}$~\cite{knight85a} states, respectively. 
The remaining intervals [Intervals (4)-(9)] , designated by dashed pink lines, correspond to intervals that can be derived from multiple combinations of two or more IR transitions wavenumbers reported in Refs.~\cite{humphreys70a,littlefield74a}.
The Interval (9) was also determined with high accuracy using laser spectroscopy~\cite{sterr95a}.
Because these IR transitions were reported for the $^{136}$Xe isotope only, the reevaluation of the first ionization energy of Xe was initially performed for the $^{136}$Xe isotope. 

In the first step of the analysis, we verified that different combinations of IR transition wavenumbers from Refs.~\cite{humphreys70a,littlefield74a} yielded the same values for Intervals (4)-(9) within the experimental uncertainties, as detailed in the supplemental material \cite{SupplementalMaterial}. These intervals can therefore not be the cause of the $\sim 1$~GHz discrepancy in the ionization energy of Xe. In the second step, we removed the Intervals (10) and (11) from the analysis and determined the positions of all levels in Fig.~\ref{Figure11} except the ionization energy from the redundant set of transitions wavenumbers connecting them in a weighted linear least squares fit, following the approach introduced by Albritton {\it et al.}~\cite{albritton76a}. The fitted level positions reproduced all transition wavenumbers within the respective experimental uncertainties. Consequently, the discrepancy in the ionization energy of Xe does not originate from the Intervals (1)-(3) either and must therefore come from a discrepancy in the ionization energies reported in Refs.~\cite{knight85a} and~\cite{herburger24a}. One should note here that the highly precise Intervals (1) (this work) and (3) (Refs.~\cite{dreissen19a,dreissen20a}) and the ionization energy (10) (Ref.~\cite{herburger24a}) were all reported after the measurement of Interval (3) (Ref.~\cite{brandi01a}) and its use in the determination of the ionization energy of Xe in Refs.~\cite{brandi01a,nist_asd_template}. The uncertainty in the ionization energy of the $(5\mathrm{p}){^5}(6\mathrm{s^{\prime}})[1/2]_{0}$ state reported in Ref.~\cite{knight85a} (0.01~cm$^{-1}$) is about one third of the $\sim1~\mathrm{GHz}$ discrepancy whereas that of the ionization energy of the $(5\mathrm{p}){^5}(6\mathrm{s})[3/2]_{2}$~state ($\sim 3$~MHz) is about 1000 times smaller \cite{herburger24a}.

\begin{table*}[tb!]
\caption{Energy intervals $\Delta E_{(i)}/(hc)$ used to determine the ${\mathrm{Xe}}^+\ (5\mathrm{p})^{5}\ ^2{\mathrm{P}}_{3/2}\leftarrow {\mathrm{Xe}}\ (5\mathrm{p})^{6} \, ^{1}\mathrm{S}_{0}$ ionization energy of $^{136}$Xe. See Fig.~\ref{Figure11} for relation between the different intervals. \label{Table3}}
\begin{ruledtabular}
\begin{tabular}{lc D{.}{.}{5.9} c}
($i$) & Interval  & \multicolumn{1}{c}{$\Delta E_{(i)}/(hc)$ (${\mathrm{cm^{-1}}}$)} & Data origin   \\
\colrule \vspace{-0.8em} \\
(1) & ($5$p)${^6}$ $^{1}$S$_{0}$ $\rightarrow$ ($5$p)${^5}$ $6$p $[1/2]_{0}$ & 80118.982918(27) & This work \\
(2) & ($5$p)${^6}$ $^{1}$S$_{0}$ $\rightarrow$ ($5$p)${^5}$ $8$s $[3/2]_{1}$ & 90932.4584217(253) & \cite{dreissen19a,dreissen20a}\\
(3) & ($5$p)${^6}$ $^{1}$S$_{0}$ $\rightarrow$ ($5$p)${^5}$ $8$d $[1/2]_{1}$ & 94228.0187(32) & \cite{brandi01a} \\
(4) & ($5$p)${^5}$ $6$p $[1/2]_{0}$ $\rightarrow$ ($5$p)${^5}$ $8$s $[3/2]_{1}$ & 10813.4721(22) & \cite{humphreys70a,littlefield74a}\footnote{For determination, see supplemental material~\cite{SupplementalMaterial}.\label{Table3_footnote1}} \\
(5) & ($5$p)${^5}$ $6$s $[3/2]_{2}$ $\rightarrow$ ($5$p)${^5}$ $8$s $[3/2]_{1}$ & 23864.8841(17) & \cite{humphreys70a,littlefield74a}\footref{Table3_footnote1} \\
(6) & ($5$p)${^5}$ $6$s $[3/2]_{2}$ $\rightarrow$ ($5$p)${^5}$ $6$p $[1/2]_{0}$ & 13051.4111(18) & \cite{humphreys70a,littlefield74a}\footref{Table3_footnote1} \\
(7) & ($5$p)${^5}$ $6$p $[1/2]_{0}$ $\rightarrow$ ($5$p)${^5}$ $8$d $[1/2]_{1}$ & 14109.0435(21)  & \cite{humphreys70a,littlefield74a}\footref{Table3_footnote1} \\
(8) & ($5$p)${^5}$ $6$s $[3/2]_{2}$ $\rightarrow$ ($5$p)${^5}$ $8$d $[1/2]_{1}$ & 27160.4546(19) & \cite{humphreys70a,littlefield74a}\footref{Table3_footnote1} \\
(9) & ($5$p)${^5}$ $6$s $[3/2]_{2}$ $\rightarrow$ ($5$p)${^5}$ $6$s$^{\prime}$~$[1/2]_{0}$ & 9129.2211(14) & \cite{humphreys70a,littlefield74a}\footref{Table3_footnote1} \\
    & & 9129.220805(51) & \cite{sterr95a} \\
(10) & ($5$p)${^5}$ $6$s $[3/2]_{2}$ $\rightarrow$ ${\mathrm{^{136}Xe}}^+\ (5\mathrm{p})^{5}\ ^2{\mathrm{P}}_{3/2}$ & 30766.207559(10) & \cite{herburger24a} \\
(11)  & ($5$p)${^5}$ $6$s$^{\prime}$~$[1/2]_{0}$ $\rightarrow$ ${\mathrm{^{136}Xe}}^+\ (5\mathrm{p})^{5}\ ^2{\mathrm{P}}_{3/2}$ & 21637.02(1) & \cite{knight85a}\footnote{This measurement was not isotopically resolved. However, the shift of the abundance-weighted value from the $^{136}$Xe value is estimated to be only $-0.00475(4)~\mathrm{cm^{-1}}$~\cite{brandi01a,jackson74a,sterr95a,herburger24a}, which is less than the uncertainty.} \\
\end{tabular}
\end{ruledtabular}
\end{table*}

In the last step of the analysis, all intervals in Table~\ref{Table4} were used to determine the positions of the levels depicted in Fig.~\ref{Figure11} in a weighted linear least-squares fit. The fit yielded a value of 97833.7800(20)~cm$^{-1}$ for the $(5\mathrm{p})^{5}\ ^2{\mathrm{P}}_{3/2}\leftarrow (5\mathrm{p})^{6}\, ^1{\rm S}_0$ ionization wavenumber of $^{136}$Xe, corresponding to the ionization energy determined using Interval (10) from Herburger {\it et al.}~\cite{herburger24a}. If Interval (10) is disregarded in the fit, a value of 97833.813(14)~cm$^{-1}$ is obtained for the ionization energy of $^{136}$Xe, which differs by $-0.033(14)$~cm$^{-1}$ ($\equiv0.99(42)~\mathrm{GHz}$) from the value determined when Interval (11) is used instead of Interval (10). We believe that to reliably determine the ionization energy by Rydberg-series extrapolation in Xe (and other rare gas atoms), it is imperative to rigorously treat channel interactions by multichannel quantum defect theory (as done in Ref.~\cite{herburger24a}). Using the Rydberg formula for the Rydberg-series extrapolation, as was done in Ref.~\cite{knight85a}, is not sufficiently accurate.
\begin{table}[tb!]
\caption{Term values (in cm$^{-1}$) of the levels of $^{136}$Xe obtained in a weighted least-squares fit based on all intervals reported in Table~\ref{Table3}. \label{Table4}}
\begin{ruledtabular}
\begin{tabular}{l D{.}{.}{5.8}}
State & \multicolumn{1}{c}{\textrm{All data}}     \\
\colrule \vspace{-0.8em} \\
$^{136}$Xe ($5$p)${^6}$ $^{1}$S$_{0}$   & 0   \\
$^{136}$Xe ($5$p)${^5}$ $6$s $[3/2]_{2}$             & 67067.5725(20)  \\
$^{136}$Xe ($5$p)${^5}$ $6$s$^{\prime}$~$[1/2]_{0}$  & 76196.7933(20)  \\
$^{136}$Xe ($5$p)${^5}$ $6$p $[1/2]_{0}$             & 80118.98292(5)  \\
$^{136}$Xe ($5$p)${^5}$ $8$s $[3/2]_{1}$             & 90932.45842(5)  \\
$^{136}$Xe ($5$p)${^5}$ $8$d $[1/2]_{1}$             & 94228.025(3)    \\
$^{136}$Xe$^+$ $(5\mathrm{p})^{5}\ ^2{\mathrm{P}}_{3/2}$ & 97833.7800(20)   \\
\end{tabular}
\end{ruledtabular}
\end{table}

\begin{table}[tb!]
\caption{Isotopic shifts $\Delta E_{\rm I}^{\mathrm{A,136}}$ of the ${\mathrm{Xe}}^+\ (5\mathrm{p})^{5}\ ^2{\mathrm{P}}_{3/2}\leftarrow { \mathrm{Xe}}\ (5\mathrm{p})^{6} \, ^{1}\mathrm{S}_{0}$ ionization energy of the $^{\rm A}$Xe isotopes with respect to that of $^{136}$Xe, determined  using data from Refs.~\cite{jackson74a,plimmer89a,borchers89a,damico99a,suzuki02a,altiere18a,herburger24a} and the present work, as detailed in the supplemental material ~\cite{SupplementalMaterial}. The second column lists the fractional natural abundances $h_{A}$ of the $^{\rm A}$Xe isotopes from Ref.~\cite{meija16a} used to calculate the abundance-weighted (AW) shift $\Delta E_{\rm I}^{\mathrm{AW,136}}$ listed in the bottom line for an isotopically unresolved measurement according to 
$\Delta E_{\rm I}^{\mathrm{AW},136} = \sum_{A} h_{A}\Delta E_{\rm I}^{A,136} /\sum_{A}h_{A}$. \label{Table5}}
\begin{ruledtabular}
\begin{tabular}{l D{.}{.}{1.10} D{.}{.}{4.6}}
Isotope A & \multicolumn{1}{c}{\textrm{$h_{A}$}} & \multicolumn{1}{c}{\textrm{{$\Delta E_{\rm I}^{\mathrm{A,136}}$ ($\mathrm{MHz}$)}}}  \\
\colrule \vspace{-0.7em} \\
$^{128}$Xe & 0.01910(13)  & -772.3(5.1) \\
$^{129}$Xe & 0.26401(138) & -716.2(3.7) \\
$^{130}$Xe & 0.04071(22)  & -574.6(1.8) \\
$^{131}$Xe & 0.21232(51)  & -559.3(8.3) \\
$^{132}$Xe & 0.26909(55)  & -397.6(1.5) \\
$^{134}$Xe & 0.10436(35)  & -224.0(2.7) \\
$^{136}$Xe & 0.08857(72)  & 0 \\
\hline \vspace{-0.7em} \\ \multicolumn{2}{l}{Abundance-weighted shift} & -477.2(2.5)
\end{tabular}
\end{ruledtabular}
\end{table}
To obtain a value of the natural-abundance-weighted ionization energy of Xe, the same procedure was repeated, based on precise values of istotopic shifts from Refs.~\cite{jackson74a,plimmer89a,borchers89a,damico99a,suzuki02a,altiere18a,herburger24a} and the present work, as detailed in the supplemental material~\cite{SupplementalMaterial}. 
The third column of Table~\ref{Table5} lists the isotopic shifts of the first-ionization energy of the $^{128}$Xe, $^{129}$Xe, $^{130}$Xe, $^{131}$Xe, $^{132}$Xe, and $^{134}$Xe isotopes relative to that of the $^{136}$Xe isotope, i.e., $\Delta E_{\rm I}^{A,136} =  E_{\mathrm{I}}(^{A}\mathrm{Xe}) - E_{\mathrm{I}}(^{136}\mathrm{Xe})$, determined from these data. From these isotopic shifts and using the natural abundances listed in the second column of Table~\ref{Table5}, the abundance-weighted isotopic shift $\Delta E_{\rm I}^{\mathrm{AW},136}$ of the ionization energy of Xe with respect to that of $^{136}$Xe was determined to be $-477.2(2.5)~\mathrm{MHz}$ ($\equiv -0.015918(83)~\mathrm{cm^{-1}}$) 
using $\Delta E_{\rm I}^{\mathrm{AW},136} = \sum_{A} h_{A}\Delta E_{\rm I}^{A,136} /\sum_{A}h_{A}$.  The natural-abundance-weighted ionization wavenumber of Xe ($E_{\rm I}/(hc)$) is thus 97833.7641(20)~cm$^{-1}$ and lies below the previously recommended value of 97833.787(11)~cm$^{-1}$~\cite{nist_asd_template} by $-0.023(11)$~cm$^{-1}$.

\section{Conclusions}
\label{Sec:Conclusions}

We have demonstrated a method for performing Doppler-free two-photon precision spectroscopy with long near-Fourier-transform-limited pulses of narrow-bandwidth UV laser radiation. The pulses were generated by amplifying pulses of NIR laser radiation cut out of the output of a single-mode cw Ti:Sa ring laser using multipass bow-tie Ti:Sa amplifiers. The amplified pulses, with duration of up to 240~ns and pulse energies of $\sim 5~\mathrm{mJ}$, were used to generate pulsed UV-laser radiation by frequency tripling in two $\beta$-barium-borate crystals. The highly nonlinear nature of the pulse-amplification and frequency-upconversion processes led to very large fluctuations in the UV pulse energies and to even larger fluctuations of the two-photon transition intensities recorded by (2+1) REMPI.
By monitoring the UV-laser intensity and the (2+1) REMPI signal for each laser shot and binning the data according to the UV pulse energies, intensity-dependent frequency shifts could be directly determined in single scans of the laser frequency. 
The procedure enabled spectral shifts from laser frequency chirps and from the ac-Stark effect to be distinguished and analyzed separately. 

We demonstrated this technique by performing measurements of the $(5\mathrm{p})^{5}6\mathrm{p} \, [1/2]_{0}\leftarrow (5\mathrm{p})^{6} \, ^{1}\mathrm{S}_{0}$ two-photon transition in Xe. The measurements revealed intensity-dependent shifts of the two-photon transition frequencies of up to $-20~\mathrm{MHz}$ resulting from frequency chirps in the Ti:Sa amplification [see Figs.~\ref{Figure5} and \ref{Figure9}(a)]. These shifts could easily have been misinterpreted as ac-Stark shifts of the two-photon transition frequencies. 
The measurements resulted in a precision determination of the $(5\mathrm{p})^{5}6\mathrm{p} \, [1/2]_{0}\leftarrow (5\mathrm{p})^{6} \, ^{1}\mathrm{S}_{0}$ two-photon transition frequencies of Xe with an absolute accuracy of 750~kHz (e.g., 80118.982918(27)~cm$^{-1}$ for $^{136}$Xe), limited by the systematic uncertainty in the determination of the residual first-order Doppler shift. Measurements of isotopic shifts resulted in uncertainties of only 200 kHz and enabled the determination of the natural-abundance-weighted two-photon transition frequency. These new results rival in accuracy the best previous measurement of a transition frequency from the ground state of Xe by Ramsey-comb spectroscopy~\cite{dreissen19a,dreissen20a}. 

Combining the results of this investigation with previous precision measurements~\cite{jackson74a,plimmer89a,borchers89a,damico99a,suzuki02a,altiere18a,herburger24a,humphreys70a,littlefield74a,knight85a,dreissen19a,dreissen20a,brandi01a,sterr95a} helped to clarify
the origin of an $\sim 1~\mathrm{GHz}$ discrepancy in the first ionization energy of Xe and led to an improved value of 97833.7800(20)~cm$^{-1}$ for the first ionization energy of $^{136}$Xe. Additionally, a new abundance-weighted first ionization energy of natural Xe of 97833.7641(20)~cm$^{-1}$ was derived, which lies below the currently recommended value of 97833.787(11)~cm$^{-1}$~\cite{nist_asd_template} by $-0.023(11)$~cm$^{-1}$.

\begin{acknowledgments}
We thank Josef A. Agner and Hansjürg Schmutz for technical support, Urs Hollenstein for experimental assistance and advice, and Eirini Toutoudaki and Holger Herburger for useful discussions. This work is supported financially by the Swiss National Science Foundation under project 200021-236716.
\end{acknowledgments}

%

\end{document}